\documentclass[final,5p,times,twocolumn]{elsarticle}

\usepackage{natbib}
\biboptions{sort&compress}
\usepackage{graphicx}
\usepackage{subcaption}
\usepackage{amstext}
\usepackage{amssymb}
\usepackage{amsmath}
\usepackage{color}
\usepackage{booktabs}
\usepackage{todonotes}
\presetkeys{todonotes}{inline}{}
\usepackage{siunitx}
\usepackage{mhchem}
\usepackage{todonotes}
\usepackage{tikz}
\usetikzlibrary{calc,angles}
\usepackage[hidelinks=true]{hyperref}
\usepackage[displaymath,switch]{lineno}

\journal{Fluid Phase Equilibria}

\renewcommand{\vec}[1]{\ensuremath{\mathbf{#1}}}

\newcommand{\pp}[1]{\ensuremath{\left( #1 \right)}}
\newcommand{\gas}{\ensuremath{\text{g}}}
\newcommand{\liquid}{\ensuremath{\ell}}
\newcommand{\solid}{\ensuremath{\text{s}}}

\newcommand{\film}{\ensuremath{\text{f}}}
\newcommand{\pore}{\ensuremath{\text{p}}}
\newcommand{\ext}{\ensuremath{\text{e}}}
\newcommand{\extl}{\ensuremath{{\text{e}_\ell}}}
\newcommand{\extr}{\ensuremath{{\text{e}_{\text{r}}}}}
\newcommand{\posl}{\ensuremath{{z_\ell}}}
\newcommand{\posr}{\ensuremath{{z_{\text{r}}}}}
\newcommand{\angl}{\ensuremath{{\theta_\ell}}}
\newcommand{\angr}{\ensuremath{{\theta_{\text{r}}}}}
\newcommand{\betal}{\ensuremath{{\beta_\ell}}}
\newcommand{\betar}{\ensuremath{{\beta_{\text{r}}}}}

\newcommand{\nuc}{\ensuremath{\text{n}}}
\newcommand{\dd}{\ensuremath{\text{d}}}
\newcommand{\transp}{\ensuremath{\text{T}}}
\newcommand{\pd}[2]{\ensuremath{\frac{\partial #1}{\partial #2}}}
\newcommand{\od}[2]{\ensuremath{\frac{\dd #1}{\dd #2}}}
\newcommand{\odd}[2]{\ensuremath{\frac{\dd^2 #1}{\dd #2 ^2}}}

\newcommand*{\dz}{\ensuremath{\Delta z}}
\newcommand*{\mz}{\ensuremath{\bar{z}}}
\newcommand*{\dRfdz}{\ensuremath{\dot{\bar{R}}^\film}}
\newcommand*{\mRf}{\ensuremath{\bar{R}^\film}}

\begin{document}

\begin{frontmatter}

\title{Thermodynamic stability of
  droplets, bubbles and thick films in open and closed pores}

\author[ntnu-ify]{Magnus Aa. Gjennestad\corref{mag-corresp}}
\ead{magnus.aa.gjennestad@ntnu.no}
\ead{magnus@aashammer.net}
\author[ntnu-ept,sintef-er]{\O{ivind} Wilhelmsen}

\cortext[mag-corresp]{Corresponding author.}

\address[ntnu-ify]{PoreLab/Department of Physics, Norwegian University
  of Science and Technology, H{\o}gskoleringen 5, NO-7491 Trondheim,
  Norway} \address[ntnu-ept]{Department of Energy and Process
  Engineering, Norwegian University of Science and Technology,
  Kolbj{\o}rn Hejes vei 1B, NO-7491 Trondheim, Norway}
\address[sintef-er]{PoreLab/SINTEF Energy Research, Sem S{\ae}lands
  vei 11, NO-7034 Trondheim, Norway}

\date{\today}

\begin{abstract}
A fluid in a pore can form diverse heterogeneous structures. We
combine a capillary description with the cubic-plus-association
equation of state to study the thermodynamic stability of droplets,
bubbles and films of water at \SI{358}{\kelvin} in a cylindrically
symmetric pore. The equilibrium structure depends strongly on the size
of the pore and whether the pore is closed (canonical ensemble) or
connected to a particle reservoir (grand canonical ensemble). A new
methodology is presented to analyze the thermodynamic stability of
films, where the integral that describes the total energy of the
system is approximated by a quadrature rule. We show that, for large
pores, the thermodynamic stability limit of adsorbed droplets and
bubbles in both open and closed pores is governed by their mechanical
stability, which is closely linked to the pore shape. This is also the
case for a film in a closed pore. In open pores, the film is
chemically unstable except for very low film-phase contact angles and
for a limited range in external pressure. This result emphasizes the
need to invoke a complete thermodynamic stability analysis, and not
restrict the discussion to mechanical stability. A common feature for
most of the heterogeneous structures examined is the appearance of
regions where the structure is metastable with respect to a pore
filled with a homogeneous fluid. In the closed pores, these regions
grow considerably in size when the pores become smaller. This can be
understood from the larger energy cost of the interfaces relative to
the energy gained from having two phases. Complete phase diagrams are
presented that compare all the investigated structures. In open pores
at equilibrium, the most stable structure is either the homogeneous
phase or adsorbed droplets and bubbles, depending on the type of phase
in the external reservoir. Smaller pores allow for droplets and bubbles
to adsorb for a larger span in pressure. In closed pores, most of the
investigated configurations can occur depending on the total density,
the contact angle and the pore shape.
The analysis presented in this work is
a step towards developing a thermodynamic framework to map the rich
heterogeneous phase diagram of porous media and other confined
systems.
\end{abstract}

\begin{keyword}
  thermodynamics \sep stability \sep droplet \sep bubble \sep film \sep pore
\end{keyword}

\end{frontmatter}


\begin{figure*}[htbp]
  \centering
  \begin{subfigure}[b]{0.5\textwidth}
    \centering
    \includegraphics[width=0.9\textwidth]{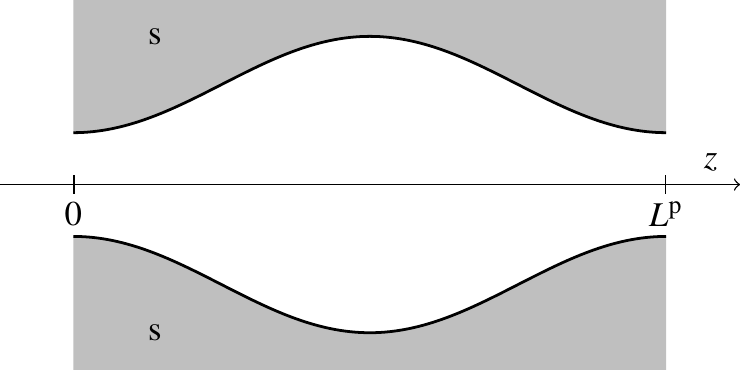}
    \caption{}
    \label{fig:pore_with_homogenous_phase}
  \end{subfigure}~
  \begin{subfigure}[b]{0.5\textwidth}
    \centering
    \includegraphics[width=0.9\textwidth]{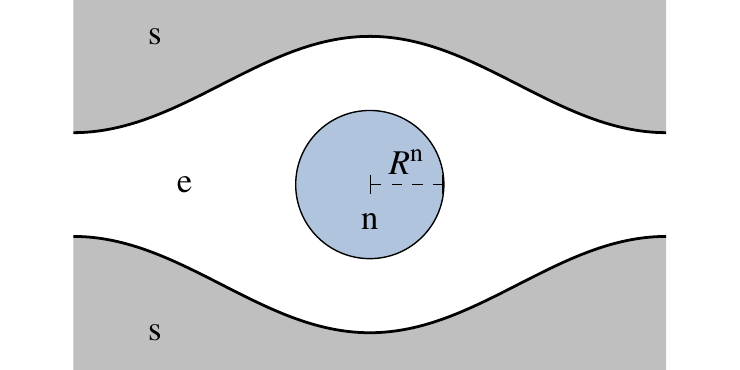}
    \caption{}
    \label{fig:pore_with_free_droplet}
  \end{subfigure}
  \begin{subfigure}[b]{0.5\textwidth}
    \centering
    \includegraphics[width=0.9\textwidth]{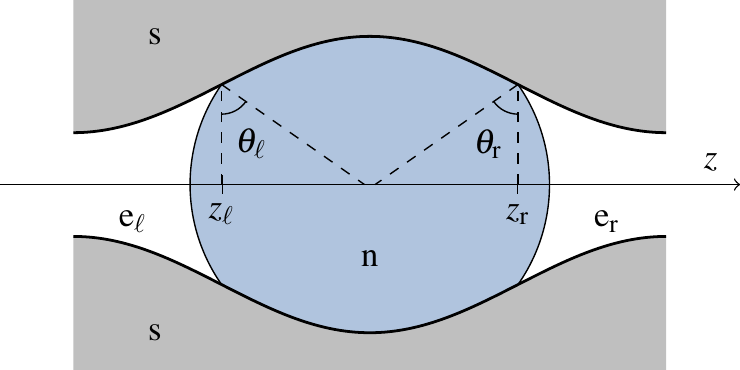}
    \caption{}
    \label{fig:pore_with_adsorbed_droplet}
  \end{subfigure}~
  \begin{subfigure}[b]{0.5\textwidth}
    \centering
    \includegraphics[width=0.9\textwidth]{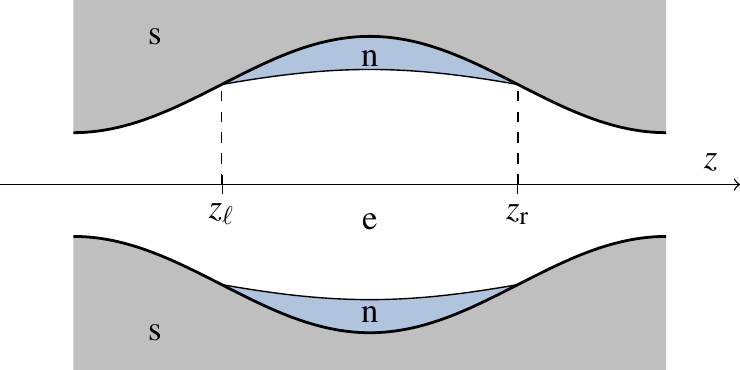}
    \caption{}
    \label{fig:pore_with_film}
  \end{subfigure}
  \caption{Illustration of the heterogeneous fluid structures under
    consideration: (a) a homogeneous fluid phase, (b) a free droplet
    or bubble that is not in contact with the pore walls, (c) a
    droplet or bubble filling the entire cross-section of some part
    pore and (d) a thick film of gas or liquid.}
  \label{fig:fluid_configurations}
\end{figure*}

\section{Introduction}
\label{sec:introduction}
Some phenomena occur exclusively in pores or under strong
confinement. In porous materials, a liquid phase can form at pressures
below the saturation pressure during capillary
condensation~\cite{Hiratsuka2017,Horikawa_2011,Neimark_2000,Neimark_2002},
liquid water can be stretched to negative pressures exceeding
\SI{140}{\mega\pascal} in quartz
inclusions~\cite{Caupin_2012,Azouzi_2013} and giant charge reversal
has been observed in confined systems filled with
electrolytes~\cite{Wang_2017}. The understanding of such systems is at
the core of widely different topics such as porous media
science~\cite{Blunt1998}, atmospheric science~\cite{Kumala_2004} and
biology~\cite{Huberman1968}.

While the thermodynamics of homogeneous systems is well
understood~\cite{Callen}, this is not the case for heterogeneous
systems, as evident e.g.\ from the large deviations between
experiments, theory and simulations for the formation of
drops~\cite{Fladerer2006,Wilhelmsen_2014}. Both in bulk systems and in
systems under confinement, equilibrium is characterized by a minimum
of an energy state function whose nature is determined by the boundary
conditions. For instance, in a closed container at constant
temperature, equilibrium is a minimum of the Helmholtz energy, while
the Gibbs energy is minimum at atmospheric conditions~\cite{Callen}

A complicating factor in pores, is that multiple heterogeneous
structures such as films, adsorbed or free droplets and bubbles, and
combinations of these, could all be stationary states of the same
energy state function~\cite{Rusanov2005}. Such states are typically
characterized by uniform temperature, equality of chemical potentials
and mechanical equilibrium~\cite{Callen,Abraham_1974}. These
conditions being satisfied however, does not imply a minimum, as the
stationary state can also be a maximum or a saddle
point~\cite{Yang1985}. To determine the equilibrium state, it is
necessary to employ thermodynamic stability
analysis~\cite{Aursand2017}, where the outcome depends strongly on the
boundary conditions. For instance, bubbles and droplets are known to
be unstable at atmospheric conditions, as they represent saddle points
in the Gibbs energy landscape~\cite{Yang1985}. However, in confined
systems, bubbles and droplets can be minima in the Helmholtz energy
and thus be stable~\cite{Wilhelmsen2014super,Wilhelmsen2015cav}.

In the literature on the stability of heterogeneous structures, many
works have studied \emph{thin} films, often in combination with
adsorbed droplets~\cite{Berg2009,Boinowich2011a,Boinowich2011b}. Films
are characterized as either thin ($\alpha$-films) or thick
($\beta$-films). In thin films, the thermodynamic properties of the
interior deviate from bulk behavior, resulting in a non-zero
disjoining pressure. Thin films have been examined by use of
theory~\cite{Neimark1999,Neimark2000}, molecular
simulations~\cite{Bhatt2002,Hu2013}, density functional
theory~\cite{Nold2014} and experiments~\cite{Checco2012}.  A common
feature of previous works in the literature discussing the stability
of films, is that they consider only stability towards perturbations
of the film height. This differs from thermodynamic stability, since
exchange of particles is neglected~\cite{Neimark1999,Neimark2000}. In
this work, we show that the thermodynamic stability of the simplest
type of film, the thick film, is very different for closed (canonical
ensemble) and open systems (grand canonical ensemble). A new
methodology to analyze the thermodynamic stability of films will be
presented. This methodology can be extended to include the disjoining
pressure and give new insight into thin films in future work.

We will discuss in detail the difference between the thermodynamic
stability of heterogeneous structures and equilibrium configurations
in \emph{open} and \emph{closed} pores, as well as the influence of
pore size. The work is a step towards developing a thermodynamic
framework to characterize heterogeneous fluid structures and
equilibrium states inside porous media.

We begin by presenting the thermodynamic description of the fluid
structures in Section~\ref{sec:models}. We employ a capillary
description, with the rationale that it gives identical results to
more sophisticated density functional theory for the thermodynamic
stability of multicomponent droplets and
bubbles~\cite{Wilhelmsen_2014}. The numerical methods are described in
Section~\ref{sec:numerical}, before results are discussed in
Section~\ref{sec:results}. Concluding remarks are provided in
Section~\ref{sec:conclusion}.

\section{Models}
\label{sec:models}
In the following, we present thermodynamic models for the four systems
illustrated in Figure~\ref{fig:fluid_configurations}. The figure
depicts a pore in an incompressible and chemically inert solid matrix.
The pore contains a single-component fluid that can have four
different configurations, (a) a homogeneous fluid phase, (b) a free
droplet or bubble that is not in contact with the pore walls, (c) a
droplet or bubble filling the entire cross-section of some part of the
pore and (d) a thick film (no disjoining pressure) of gas or
liquid. Thin films, that are influenced by a disjoining pressure, are
beyond the scope of the present work. By deriving thermodynamic models
for all these sub-systems with the same assumptions, it is possible to
evaluate their local stability, compare their energies and thus
identify the appropriate equilibrium configuration, at given
conditions.

We consider a cylindrically symmetric pore of length $L^\pore$ and
radius $R^\pore$. The radius depends on the axial coordinate $z \in
\left[0, L^\pore\right]$. Figure~\ref{fig:pore_with_homogenous_phase}
illustrates one possible pore geometry, but the governing equations
will be derived for an arbitrary function $R^\pore(z)$, which is
sufficiently smooth for $\dot{R}^\pore = \dd R^\pore / \dd z$ and
$\ddot{R}^\pore = \dd^2 R^\pore / \dd z^2$ to be defined.

For simplicity, we will restrict the pore radii considered in the
present work to functions on the form,
\begin{linenomath} \begin{align}
  \label{eq:R_p}
  R^\pore \left( z \right) &= L^\pore \left(0.2 - 0.075 \left\{ 1 +
  \cos \left(\frac{2 \pi z}{L^\pore}\right) \right\} \right),
\end{align} \end{linenomath}
where Figure~\ref{fig:fluid_configurations} shows an example of this
profile. It is assumed that the solid matrix acts as a thermal
reservoir for the fluids at temperature $T$.

With the above assumptions, the volume of the pore is constant and can
be calculated from,
\begin{linenomath} \begin{align}
  \label{eq:V_p}
  V^\pore &= \pi \int_0^{L^\pore} \pp{ R^\pore }^2 \dd z.
\end{align} \end{linenomath}
Similarly, the surface area of the solid matrix is constant and equal
to
\begin{linenomath} \begin{align}
  \label{eq:A_p}
  A^\pore &= 2 \pi \int_0^{L^\pore} R^\pore \sqrt{1 +
    \pp{\dot{R^\pore}}^2} \dd z.
\end{align} \end{linenomath}

Three interfacial tensions will be parameters in our models. These are
$\sigma^{\gas\solid}$, $\sigma^{\liquid\solid}$ and
$\sigma^{\gas\liquid}$ for the gas-solid, liquid-solid and gas-liquid
interfaces, respectively. In terms of the interfacial tensions,
Young's law gives the contact angle $\alpha$ (measured in the liquid)
as
\begin{linenomath} \begin{align}
  \cos \pp{\alpha} &= \frac{\sigma^{\gas\solid} -
    \sigma^{\liquid\solid}}{\sigma^{\gas\liquid}}.
  \label{eq:yl}
\end{align} \end{linenomath}
Due to the isothermal conditions, interfacial tensions are assumed to
be constant.

The thermodynamic properties of the fluids are described by an
equation of state (EOS), where any EOS capable of describing the
liquid and gas phases is applicable.

In the following, we present the governing equations for all the
sub-systems in Figure~\ref{fig:fluid_configurations}.  A clear
distinction is made between whether the system is closed (the canonical
ensemble), or connected to a particle reservoir (grand canonical
ensemble). The grand canonical ensemble is a natural representation of
an open pore, which is equivalent to a system connected to a
temperature and pressure reservoir for a single-component system due
to the Gibbs--Duhem relation (see \cite{Wilhelmsen2015cav} for a
discussion).  Equilibrium in the canonical ensemble is a minimum of
the total Helmholtz energy of the system, while equilibrium in the
grand canonical ensemble is a minimum of the total grand potential
energy. These energy state functions and their stationary states have
to be identified for each of the configurations in
Figure~\ref{fig:fluid_configurations}.

\subsection{Pore with a homogeneous phase}
We start by considering the simplest fluid configuration possible,
which is a pore filled with a single-phase fluid. This type of
configuration is illustrated in
Figure~\ref{fig:pore_with_homogenous_phase}. The Helmholtz energy
of this system is
\begin{linenomath} \begin{align}
  F &= -p V^\pore + \mu N + \sigma A^\pore,
\end{align} \end{linenomath}
where $p$ is the fluid pressure, $\mu$ is the chemical potential, $N$
is the number of particles and $\sigma$ is the interfacial tension
between the solid and the fluid, i.e.\ $\sigma^{\gas\solid}$ for a
  pore filled with gas and $\sigma^{\liquid\solid}$ for a pore filled
  with liquid. The grand potential energy is
\begin{linenomath} \begin{align}
  \label{eq:Omega_single-phase}
  \Omega &= F - \mu N.
\end{align} \end{linenomath}
Since the solid matrix is incompressible, chemically inert and has a
constant temperature, its Helmholtz and grand potential energies are
constants that can be omitted from the analysis without any further
effect on the results. A stationary state of a homogeneous phase is
characterized by uniform temperature, pressure and chemical
potentials~\cite{Callen}.  The phase is thermodynamically stable if
its density is within the spinodal limits at the specified
temperature. We refer to \citet{Aursand2017} for further details on
how the spinodal limits can be calculated.

\subsection{Pore with a free droplet or bubble}
\label{sec:pfdb}
Next, we consider a pore with a free spherical droplet or bubble that
is not in contact with the pore walls. The droplet/bubble phase is
labeled \nuc, while the surrounding phase is labeled $\ext$, as
illustrated in Figure~\ref{fig:pore_with_free_droplet}.

First, we assume that the pore is a closed system with a fixed total
number of particles $N$, total volume $V$ and temperature $T$. An
equilibrium state of this system is then a minimum in the total
Helmholtz energy of the system,
\begin{linenomath} \begin{align}
  F =& -p^\ext V^\ext + \mu^\ext N^\ext + \sigma^{\ext\solid}
  A^{\ext\solid} - p^\nuc V^\nuc + \mu^\nuc N^\nuc + \sigma^{\ext\nuc}
  A^{\ext\nuc}.
\end{align} \end{linenomath}
Herein, $p^j$ is the pressure of phase $j$, $\mu^j$ is the chemical
potential of phase $j$ and $N^j$ is the number of particles in phase
$j$. The area of the interface between phase $i$ and $j$ is denoted by
$A^{ij}$ and the tension of this interface by $\sigma^{ij}$. Using
that $V^\ext + V^\nuc = V^\pore$, $N^\ext + N^\nuc = N$,
$A^{\ext\solid} = A^\pore$ and the Gibbs--Duhem relations for each
phase, the differential of $F$ can be written as
\begin{linenomath} \begin{align}
  \dd F =& -\left( p^\nuc - p^\ext \right) \dd V^\nuc + \left(
  \mu^\nuc - \mu^\ext \right) \dd N^\nuc + \sigma^{\ext\nuc} \dd
  A^{\ext\nuc}.
\end{align} \end{linenomath}
Since the \nuc-phase is assumed to be spherical, $A^{\ext\nuc}$ and
$V^\nuc$ are not independent. We choose to describe the geometry of
the droplet/bubble in terms of its radius $R^\nuc$ and get the
differential
\begin{linenomath} \begin{align}
  \dd F =& -\left( p^\nuc - p^\ext - \frac{2 \sigma^{\ext\nuc}}
      {R^\nuc} \right) 4 \pi \pp{R^\nuc}^2 \dd R^\nuc + \left(
      \mu^\nuc - \mu^\ext \right) \dd N^\nuc,
\end{align} \end{linenomath}
in terms of perturbations in the independent free variables of the
system, $R^\nuc$ and $N^\nuc$. The elements in the Jacobian vector of
$F$ are then
\begin{linenomath} \begin{align}
  \label{eq:dFdR_n_free_droplet}
  \pp{\pd{F}{R^\nuc}}_{N^\nuc} &= -4 \pi \pp{R^\nuc}^2
  \left(p^\nuc - p^\ext - \frac{2 \sigma^{\ext\nuc}} {R^\nuc} \right),
\end{align} \end{linenomath}
and
\begin{linenomath} \begin{align}
  \label{eq:dFdN_n_free_droplet}
  \pp{\pd{F}{N^\nuc}}_{R^\nuc} &= \mu^\nuc - \mu^\ext.
\end{align} \end{linenomath}
A stationary state of $F$ is therefore characterized by equality of
the chemical potentials in the two phases and a pressure difference
between the gas and liquid given by the Young--Laplace equation. The
Hessian matrix of $F$ can be found by further differentiation of the
Jacobian, as shown in \cite{Wilhelmsen_2014}.

Let us now consider the pore in an open system at fixed total volume
and temperature that is connected to a particle reservoir, such that
the chemical potential of the $\ext$-phase is fixed. An equilibrium
state is then a minimum of the total grand potential energy of the
system,
\begin{linenomath} \begin{align}
  \label{eq:Omega_two_phase}
  \Omega = F - \mu^\ext N.
\end{align} \end{linenomath}
By a derivation analogous to that above, one finds the Jacobian vector
of $\Omega$ and that the criteria for a stationary point in the grand
canonical ensemble are exactly the same as in the canonical
ensemble. One subtle difference that makes the Hessian matrix of
$\Omega$ different from that of $F$, is that $\mu^\ext$ in the open
system no longer depends on the free variables of the system.

\subsection{Pore with an adsorbed droplet or bubble}
The next fluid configuration we consider is a pore containing a
droplet or a bubble that is in contact with the pore walls and fills
the entire pore cross-section for some interval on the $z$-axis. This
type of fluid configuration is illustrated in
Figure~\ref{fig:pore_with_adsorbed_droplet}. The two fluid-fluid
interfaces, in contact with the pore walls at $\posl$ and $\posr$, are
assumed to be spherical caps. Again, the droplet/bubble phase is
labeled $\nuc$. The fluid phase on the left side is labeled $\extl$
and the one on the right side is labeled $\extr$.

The Helmholtz energy is
\begin{linenomath} \begin{align}
  F =& \nonumber -p^\extl V^\extl + \mu^\extl N^\extl +
  \sigma^{\ext\solid} A^{\extl\solid} + \sigma^{\ext\nuc}
  A^{\extl\nuc} \\ \nonumber &- p^\extr V^\extr +\mu^\extr N^\extr +
  \sigma^{\ext\solid} A^{\extr\solid} + \sigma^{\ext\nuc}
  A^{\extr\nuc} \\ &- p^\nuc V^\nuc + \mu^\nuc N^\nuc +
  \sigma^{\nuc\solid} A^{\nuc\solid}.
  \label{eq:F_adsorbed_droplet}
\end{align} \end{linenomath}
Using now that $V^\extl + V^\extr + V^\nuc = V^\pore$,
$A^{\extl\solid} + A^{\nuc\solid} + A^{\extr\solid} = A^\pore$,
$N^\extl + N^\nuc + N^\extr = N$ and that the Gibbs--Duhem relation is
satisfied for each phase, we can formulate the differential of $F$ as
\begin{linenomath} \begin{align}
  \dd F =& \nonumber -\left(p^\extl - p^\nuc \right) \dd V^\extl +
  \left(\mu^\extl - \mu^\nuc \right) \dd N^\extl \\ \nonumber &+
  \left( \sigma^{\ext\solid} - \sigma^{\nuc\solid} \right) \dd
  A^{\extl\solid} + \sigma^{\ext\nuc} \dd A^{\extl\nuc} \\ \nonumber
  &- \left( p^\extr - p^\nuc \right) \dd V^\extr + \left( \mu^\extr -
  \mu^\nuc \right) \dd N^\extr \\ &+ \left( \sigma^{\ext\solid} -
  \sigma^{\nuc\solid} \right) \dd A^{\extr\solid} + \sigma^{\ext\nuc}
  \dd A^{\extr\nuc}.
\end{align} \end{linenomath}
Since both fluid-fluid interfaces are assumed to be shaped like
spherical caps, they can each be described by two independent
variables. We therefore parameterize the six geometrical quantities
$V^\extl$, $A^{\extl\solid}$, $A^{\extl\nuc}$, $V^\extr$,
$A^{\extr\solid}$, $A^{\extr\nuc}$ in terms of the four independent
variables $\posl$, $\posr$, $\angl$, and $\angr$. As illustrated in
Figure~\ref{fig:pore_with_adsorbed_droplet}, $\posl$ denotes the
position along the $z$-axis of the left three-phase contact line of
the left meniscus and $\posr$ denotes the position of the contact line
of the right meniscus. The angle $\angl$ is between a line connecting
the center of the left sphere with a point on the left three-phase
contact line and a line from the same point on the contact line which
is perpendicular to the $z$-axis. The angle $\angr$ is defined
analogously. In terms of the independent variables, we have that
\begin{linenomath} \begin{align}
  V^{\extl} &= \pi \int_0^\posl \pp{R^\pore}^2 \dd z - \zeta
  \pp{\posl, \angl}, \label{eq:Adrop} \\ V^{\extr} &= \pi \int_\posr^{L^\pore}
  \pp{R^\pore}^2 \dd z - \zeta \pp{\posr, \angr}, \\ A^{\extl\solid}
  &= 2 \pi \int_0^\posl R^\pore \sqrt{ 1 + \pp{\dot{R^\pore}}^2 }\dd
  z, \\ A^{\extr\solid} &= 2 \pi \int_\posr^{L^\pore} R^\pore \sqrt{ 1
    + \pp{\dot{R^\pore}}^2 }\dd z ,\\ A^{\extl\nuc} &= \pi \left\{
  R^\pore \pp{\posl} \right\}^2 \left\{ 1 + \xi^2 \pp{\angl} \right\},
  \\ A^{\extr\nuc} &= \pi \left\{ R^\pore \pp{\posr} \right\}^2
  \left\{ 1 + \xi^2 \pp{\angr} \right\},
\end{align} \end{linenomath}
where
\begin{linenomath} \begin{align}
  \zeta \pp{z, \theta} &= \frac{\pi}{6} \left\{R^\pore
  \pp{z}\right\}^3 \xi \pp{\theta} \left\{3 + \xi^2 \pp{\theta}
  \right\}, \\ \xi \left( \theta \right) &= \frac{1 - \sin
    \pp{\theta}}{\cos \pp{\theta}}.
\end{align} \end{linenomath}
The Helmholtz energy differential may then be expressed as
\begin{linenomath} \begin{align}
  \dd F =& \nonumber \left\{ \mu^\extl - \mu^\nuc \right\} \dd N^\extl
  + \left\{ \mu^\extr - \mu^\nuc \right\} \dd N^\extr \\ \nonumber &+
  \left\{ \left(\sigma^{\ext\solid} - \sigma^{\nuc\solid} \right)
  \pd{A^{\extl\solid}}{\posl} + \sigma^{\ext\nuc}
  \pd{A^{\extl\nuc}}{\posl} - \left( p^\extl - p^\nuc \right)
  \pd{V^\extl}{\posl} \right\} \dd \posl \\ \nonumber &+ \left\{
  \left(\sigma^{\ext\solid} - \sigma^{\nuc\solid} \right)
  \pd{A^{\extr\solid}}{\posr} + \sigma^{\ext\nuc}
  \pd{A^{\extr\nuc}}{\posr} - \left( p^\extr - p^\nuc \right)
  \pd{V^\extr}{\posr} \right\} \dd \posr \\ \nonumber &+ \left\{
  \sigma^{\ext\nuc} \pd{A^{\extl\nuc}}{\angl} - \left( p^\extl -
  p^\nuc \right) \pd{V^\extl}{\angl} \right\} \dd \angl \\ &+ \left\{
  \sigma^{\ext\nuc} \pd{A^{\extr\nuc}}{\angl} - \left( p^\extr -
  p^\nuc \right) \pd{V^\extr}{\angr} \right\} \dd \angr.
  \label{eq:dF_adsorbed_droplet}
\end{align} \end{linenomath}
Here, the expressions in the curly brackets are the elements of the
Jacobian vector of $F$. The Hessian matrix can be found by further
differentiation of the Jacobian vector.

In a stationary state, all the terms of~\eqref{eq:dF_adsorbed_droplet}
must vanish. Setting the last two terms equal to zero, yields
\begin{linenomath} \begin{align}
  \label{eq:dpl_adsorbed_droplet}
  p^\nuc - p^\extl &= \frac{2 \sigma^{\ext\nuc} \cos
    \pp{\angl}}{R^\pore \pp{\posl}},
  \\ \label{eq:dpr_adsorbed_droplet} p^\nuc - p^\extr &= \frac{2
    \sigma^{\ext\nuc} \cos \pp{\angr}}{R^\pore \pp{\posr}}.
\end{align} \end{linenomath}
Since the radii of curvature of the fluid-fluid interfaces are
$R^\pore \pp{\posl}/\cos \pp{\angl}$ and $R^\pore \pp{\posr}/\cos
\pp{\angr}$, these two equations imply that the interfaces obey the
Young--Laplace equation.

For the first two terms in equation~\eqref{eq:dF_adsorbed_droplet} to
vanish, we must have equality of the chemical potential in all fluid
phases. This requires that $p^\extl = p^\extr$, and
\eqref{eq:dpl_adsorbed_droplet} and \eqref{eq:dpr_adsorbed_droplet}
may then be combined to give
\begin{linenomath} \begin{align}
  \label{eq:equal_k_adsorbed_droplet}
  \frac{\cos \pp{\angl}}{R^\pore \pp{\posl}} &= \frac{\cos
    \pp{\angr}}{R^\pore \pp{\posr}},
\end{align} \end{linenomath}
meaning that both fluid-fluid interfaces must have the same curvature.
The relation between $\theta_i$ and the contact angle, measured in
the $\nuc$-phase, is
\begin{linenomath} \begin{align}
  \alpha^\nuc &= \theta_i + \beta_i,
\end{align} \end{linenomath}
where 
\begin{linenomath} \begin{align}
  \betal &= \arctan \left( \dot{R^\pore} \pp{\posl} \right), \\ \betar
  &= \arctan\left(-\dot{R^\pore} \pp{\posl}\right).
\end{align} \end{linenomath}
By combining the above equations with
\eqref{eq:Adrop}-\eqref{eq:dF_adsorbed_droplet} we find by use of
trigonometric relations that both contact angles obey Young's equation
\eqref{eq:yl} in a stationary state.

The grand canonical energy $\Omega$ of the system is given by
\eqref{eq:Omega_two_phase}, where \eqref{eq:F_adsorbed_droplet} is
used for the Helmholtz energy. The derivatives of $\Omega$ may then be
found by an analogous derivation to that given above. One result from
this derivation is that the criteria for a stationary state of
$\Omega$ are the same as those for a stationary state of $F$. The
Hessian matrix differs, however.

\subsection{Pore with a thick film of liquid or gas}
\label{sec:tfilm}
The final fluid configuration that will be considered is a pore with a
wetting film consisting of either liquid or vapor. The film is
considered so thick that interactions between the fluid-fluid and
fluid-solid interfaces, as modeled by the disjoining pressure, are
negligible. We refer to excellent works in the literature for further
information about the disjoining
pressure~\cite{Berg2009,Boinowich2011a,Boinowich2011b}. The thick film
configuration is illustrated in Figure~\ref{fig:pore_with_film}. As
for the adsorbed droplet and bubble, $\posl$ and $\posr$ denote the
positions of the left and right three-phase contact lines,
respectively.

The Helmholtz energy is now
\begin{linenomath} \begin{align}
  F =& \nonumber -p^\ext V^\ext + \mu^\ext N^\ext +
  \sigma^{\ext\solid} A^{\ext\solid} + \sigma^{\ext\nuc} A^{\ext\nuc}
  \\ &- p^\nuc V^\nuc + \mu^\nuc N^\nuc + \sigma^{\nuc\solid}
  A^{\nuc\solid}.
  \label{eq:F_full_film}
\end{align} \end{linenomath}
The interfacial area between the $\nuc$- and $\solid$-phases is
a function of $\posl$ and $\posr$,
\begin{linenomath} \begin{align}
  \label{eq:A_ns_film}
  A^{\nuc\solid} &= 2 \pi \int_\posl^\posr R^\pore \sqrt{ 1 +
    \pp{\dot{R^\pore}}^2 } \dd z.
\end{align} \end{linenomath}
The volume of the $\nuc$-phase $V^\nuc$ and the interfacial area
between the $\nuc$- and $\ext$-phases $A^{\ext\nuc}$ depend on
$\posl$ and $\posr$ and on the shape of the fluid-fluid interface in
between. Since the system is axisymmetric, we may express the shape of
the interface by the function $R^\film\pp{z}$, which represents the
distance between a point on the fluid-fluid interface to its closest
point on the $z$-axis. The volume $V^\nuc$ and area $A^{\ext\nuc}$ are
then \emph{functionals} of $R^\film$,
\begin{linenomath} \begin{align}
  \label{eq:A_en_film}
  A^{\ext\nuc} &= \int_\posl^\posr L_{A^{\ext\nuc}} \pp{z, R^\film,
    \dot{R}^\film} \dd z, \\ \label{eq:V_n_film} V^\nuc &=
  \int_\posl^\posr L_{V^\nuc} \pp{z, R^\film, \dot{R}^\film} \dd z.
\end{align} \end{linenomath}
The integrands of these functionals are
\begin{linenomath} \begin{align}
  \label{eq:L_Aen} L_{A^{\ext\nuc}} \pp{z, R^\film,
    \dot{R}^\film} &= 2 \pi R^\film \sqrt{1 + \pp{\dot{R}^\film}^2},
  \\ \label{eq:L_Vn} L_{V^\nuc} \pp{z, R^\film, \dot{R}^\film} &= \pi
  \left\{\pp{R^\pore}^2 - \pp{R^\film}^2 \right\}.
\end{align} \end{linenomath}
Using that $A^{\ext\solid} + A^{\nuc\solid} = A^\pore$, $V^\ext +
V^\nuc = V^\pore$ and $N^\ext + N^\nuc = N$ we proceed to eliminate
$A^{\ext\solid}$, $V^\ext$ and $N^\ext$ from \eqref{eq:F_full_film}
and get
\begin{linenomath} \begin{align}
  F =\ & \nonumber \left(\mu^\nuc - \mu^\ext \right) N^\nuc - p^\ext
  V^\pore + \sigma^{\ext\solid} A^\pore + \mu^\ext N \\ & - \left(
  p^\nuc - p^\ext \right) V^\nuc + \left( \sigma^{\nuc\solid} -
  \sigma^{\ext\solid} \right) A^{\nuc\solid} + \sigma^{\ext\nuc}
  A^{\ext\nuc}.
  \label{eq:F_full_film_eliminated}
\end{align} \end{linenomath}
Taking the differential on both sides of
\eqref{eq:F_full_film_eliminated} gives
\begin{linenomath} \begin{align}
dF =\ & \nonumber
\left(\mu^\nuc - \mu^\ext \right) dN^\nuc  - \left( p^\nuc - p^\ext
\right) dV^\nuc \\ &+ \left( \sigma^{\nuc\solid} - \sigma^{\ext\solid}
\right) dA^{\nuc\solid} + \sigma^{\ext\nuc} dA^{\ext\nuc}.
 \label{eq:F_full_film_eliminated_diff}
\end{align} \end{linenomath}
The above equation shows that in a stationary state, the chemical
potentials of the $\ext$- and $\nuc$-phases must be the same. The
Helmholtz energy of a film in a pore with a uniform chemical potential
(subscript $\mu$) can be formulated as
\begin{linenomath} \begin{align}
  F_\mu =\ & \nonumber F_\mu^\circ + \pi \int_\posl^\posr 2
  \sigma^{\ext\nuc} R^\film \sqrt{1 + \pp{\dot{R}^\film}^2}
  \\ \nonumber & \quad - \left( p^\nuc - p^\ext \right)
  \left\{\pp{R^\pore}^2 - \pp{R^\film}^2 \right\} \\ & \quad + 2
  \left( \sigma^{\nuc\solid} - \sigma^{\ext\solid} \right) R^\pore
  \sqrt{ 1 + \pp{\dot{R^\pore}}^2 } \dd z.
  \label{eq:film}
\end{align} \end{linenomath}
where we have integrated \eqref{eq:F_full_film_eliminated_diff} and
used \eqref{eq:A_en_film} and \eqref{eq:V_n_film}. $F_{\mu}^\circ$ is
a constant. To have a stationary state in $F_{\mu}$, the first
variation with respect to the function $R^\film$ must vanish and
$R^\film$ must therefore satisfy the Euler--Lagrange equation in the
interval $z \in \pp{\posl, \posr}$. The Euler--Lagrange equation leads
to the following second-order ordinary differential equation (ODE) for
$R^\film$,
\begin{linenomath} \begin{align}
  \label{eq:film_ode}
   p^\ext - p^\nuc &= \sigma^{\ext\nuc}\left( \kappa_1 + \kappa_2
   \right),
\end{align} \end{linenomath}
where
\begin{linenomath} \begin{align}
  \kappa_1 &= \frac{1}{R^\film \left( 1 + \pp{\dot{R}^\film}^2
    \right)^{\frac{1}{2}}}, \\ \kappa_2 &= -
  \frac{\ddot{R}^\film}{\left(1 +
    \pp{\dot{R}^\film}^2\right)^{\frac{3}{2}}},
\end{align} \end{linenomath}
are the interfacial curvatures. This ODE can be recognized as the
Young--Laplace relation for the film. Since \eqref{eq:film_ode} is a
second-order ODE, we need boundary conditions on both $R^\film$
and $\dot{R}^\film$ at the free end points, $\posl$ and $\posr$ to
fully define the film. The boundary conditions on $R^\film$ are
$R^\film \pp{\posl} = R^\pore \pp{\posl}$ and $R^\film \pp{\posr} =
R^\pore \pp{\posr}$. To derive boundary conditions for
$\dot{R}^\film$, we must consider the \textit{transversal conditions}
at the free end points $\posl$ and $\posr$, see e.g.\ page 159 in
\cite{Troutman1996}. They give that
\begin{linenomath} \begin{align}
  \label{eq:film_contact_angle}
  \cos \pp{\alpha^\nuc} &= \frac{\sigma^{\ext\solid} -
    \sigma^{\nuc\solid}}{\sigma^{\ext\nuc}},
\end{align} \end{linenomath}
must be satisfied at the end points $\posl$ and $\posr$. Here,
$\alpha^\nuc$ is the contact angle measured in the film. The
transversal conditions are thus satisfied when the three-phase contact
angles obey Young's equation, like the adsorbed droplet/bubble.

The grand canonical energy $\Omega$ of the pore with a film is given
by \eqref{eq:Omega_two_phase}, where \eqref{eq:F_full_film} is used
for the Helmholtz energy. As for the other systems, the criteria for a
stationary state of $\Omega$ are the same as those for a stationary
state of $F$. To analyze the thermodynamic stability of the film, one
possibility is to study the second variation of e.g.\ $F$ at the
stationary state. As this is often challenging due to the infinite
number of possible functions that can perturb the stationary state, we
present in Section~\ref{sec:numerical} a new methodology to analyze
the stationary states of films.

\section{Numerical methods}
\label{sec:numerical}
In this section, we provide details on the numerical methods used to
determine the stationary states and the thermodynamic stability of the
configurations in Figure~\ref{fig:fluid_configurations}. For all the
heterogeneous structures, one way to identify stationary states is to
first determine the shape and positions of the interfaces. The outcome
is a fixed value of the pressure difference $\Delta p = p^\ext -
p^\nuc$. For the film, this can be done by solving the Euler--Lagrange
equation \eqref{eq:film_ode} as described in
Section~\ref{sec:numfilm_ode}. Subsequently, one can calculate the
phase equilibrium with specified $\Delta p = p^\ext - p^\nuc$, as
described in Section~\ref{sec:phase_eq}.

\subsection{Phase equilibrium calculations}
\label{sec:phase_eq}
All the heterogeneous structures considered in this work are
characterized by the same chemical potentials in the $\nuc$- and
$\ext$-phases, but at a fixed pressure difference $\Delta p = p^\ext -
p^\nuc$, temperature $T$ and phase volumes $V^\ext$ and $V^\nuc$. This
poses an untypical phase-equilibrium problem.

To determine the remaining thermodynamic properties of the system, one
can solve for the dimensionless particle numbers $N^\nuc/N^\circ$ and
$N^\ext/N^\circ$. This procedure amounts to solving the non-linear
system of equations $\vec{F} \left( N^\nuc/N^\circ, N^\ext/N^\circ
\right) = \vec{0}$, where
\begin{linenomath} \begin{align}
  \label{eq:phase_eq}
  \vec{F} \left( \frac{N^\nuc}{N^\circ}, \frac{N^\ext}{N^\circ}
  \right) =
  \begin{bmatrix}
    \frac{ \mu^\ext \pp{T, V^\ext, N^\ext} - \mu^\nuc \pp{T, V^\ext,
        N^\ext}}{RT} \\ \frac{ p^\ext \pp{T, V^\ext, N^\ext} - p^\nuc
      \pp{T, V^\ext, N^\ext} - \Delta p }{p^\circ}
  \end{bmatrix}.
\end{align} \end{linenomath}
Herein, the functions for pressure and chemical potentials, and the
derivatives required to compute the Jacobian matrix of $\vec{F},$ are
provided by the EOS. The scaling parameters are
\begin{linenomath} \begin{align}
  p^\circ &= \SI{e5}{\pascal}, \\
  N^\circ &= \frac{p^\circ V^\pore}{RT},
\end{align} \end{linenomath}
and $R$ is the universal gas constant.  The system in
\eqref{eq:phase_eq} was solved using Newton's method.  Initial guesses
for $N^\nuc$ and $N^\ext$ were obtained from a standard phase
equilibrium calculation~
\cite{Wilhelmsen2013_TDAE,Aasen2017,Mollerup_book,Aursand2017} at the
specified temperature and saturation pressure. The EOS implementation
used was provided by our in-house thermodynamic library presented by
\citet{Wilhelmsen2017}.

\subsection{Solving the film Euler--Lagrange equation}
\label{sec:numfilm_ode}
The ODE in~\eqref{eq:film_ode} gives a requirement on the film profile
that must be satisfied to have a vanishing first variation of the
Helmholtz and grand canonical energies. Since the ODE is second-order
and requires boundary conditions on both $R^\film$ and $\dot{R}^\film$
at $\posl$ and $\posr$, it represents a two-point boundary value
problem. We solved this problem using the shooting method. The
solution strategy was to first specify the position $\posl$. Since
the contact angle and pore radius at $\posl$ are known, $R^\film$ and
$\dot{R}^\film$ are also specified. Next, a search was performed for
the values of the variables $\Delta p = p^\ext - p^\nuc$ and $\posr$
that satisfy the two boundary conditions on $R^\film$ and
$\dot{R}^\film$ at $\posr$. The shooting procedure thus amounts to
solving $\vec{G} \left( \Delta p/p^\circ, \posr/L^\pore \right) = \vec{0}$, where
\begin{linenomath} \begin{align}
  \vec{G} \left( \frac{\Delta p}{p^\circ}, \frac{\posr}{L^\pore} \right) &=
  \begin{bmatrix}
    \frac{R^\film \left( \posr \right) - R^\pore \left( \posr \right)}{L^\pore} \\
    \frac{\dot{R}^\film \left( \posr \right) -
    \tan \left( \angr \right)}{\max\left(1, |\tan \left( \angr
    \right)| \right)} \\
  \end{bmatrix},
  \label{eq:shooting_residuals}
\end{align} \end{linenomath}
and
\begin{linenomath} \begin{align}
  \angr &= \arctan \pp{\dot{R}^\pore \pp{\posr}} + \alpha^\nuc.
\end{align} \end{linenomath}
The scaling parameter is
\begin{linenomath} \begin{align}
  p^\circ &= \frac{\sigma^{\ext\nuc}}{L^\pore}.
\end{align} \end{linenomath}
One evaluation of $\vec{G}$ involves one integration of
\eqref{eq:film_ode}. We used \verb|odeint| from \verb|scipy|'s
\verb|integrate| module for the ODE integrations and \verb|fsolve|
from the \verb|optimize| module to solve $\vec{G} = \vec{0}$
\cite{SciPy}.

A complicating factor in the search for stationary states is that
there may be many solutions to $\vec{G} = \vec{0}$ with the same
$\posl$. In practice, however, we have found that we can identify
the one that is potentially stable and discard any other solutions
in subsequent analysis. This is further explained in
\ref{app:film_modes}.

\subsection{A discrete method for describing the film}
\label{sec:film_discr}
The variational formulation works well for identifying stationary
states in $F$ and $\Omega$, where the Euler--Lagrange equation for the
film~\eqref{eq:film_ode} can be solved as described in
Section~\ref{sec:numfilm_ode}. The procedure identifies stationary
states, but it does not give any insight into the thermodynamic
stability of the film. This information is contained in the second
variation (or higher-order variations, if the second variation happens
to be zero). For a stationary state of a functional to be a minimum,
it is necessary to have a positive second variation for all viable
perturbations, as discussed by~\citet{Wilhelmsen_2014}. To establish
that this is the case, or not, can be very demanding and it is
impossible for many examples.

In this section, we present a new methodology for analyzing the
thermodynamic stability of films. The approach that we follow here is
to discretize the functionals for $A^{\ext\nuc}$ \eqref{eq:A_en_film}
and $V^\nuc$ \eqref{eq:V_n_film} and use the discretized functionals
to represent the Helmholtz and grand canonical energies. The
functionals are integrated numerically using a quadrature rule over a
predefined grid, where the end-points are left unspecified. This
transforms the variational problem of minimizing $F$ or $\Omega$ in
the space of functions $R^\film$, to an algebraic problem where $F$ is
to be minimized by a vector in $\mathbb{R}^{M}$. Local stability can
then be evaluated by considering the eigenvalues of a Hessian matrix.

This procedure can be applied to general problems in functional
optimization. We have tested it carefully and successfully reproduced
well-known results from variational calculus, such as the
Brachistochrone and the hanging cable problems~\cite{Troutman1996},
see supplementary material.

In the discrete formulation of the film, we approximate the function
$R^\film$ by the vector
\begin{linenomath} \begin{align}
  \vec{R}^\film &= \left[ R^\film_1, R^\film_2, ..., R^\film_M \right]^\transp,
\end{align} \end{linenomath}
which represents the values of $R^\film$ at points on a predefined
gird with $M$ points on the $z$-axis, given by
\begin{linenomath} \begin{align}
  \vec{z} = \left[z_1, z_2, ..., z_M \right]^\transp.
\end{align} \end{linenomath}
The complete vector of geometrical unknowns, including the positions
of the free end points, is then
\begin{linenomath} \begin{align}
  \vec{x} &= \left[ x_1, x_2, ..., x_{M+2} \right]^\transp, \\ &=
  \left[ \posl, \posr, R^\film_1, R^\film_2, ..., R^\film_M
    \right]^\transp.
\end{align} \end{linenomath}
The volume $V^\nuc$ and area $A^{\nuc\ext}$ can now be approximated by
the midpoint rule,
\begin{linenomath} \begin{align}
  V^\nuc \left( \vec{x} \right) &= \sum_{i=0}^M L_{V^\nuc} \left(
  \mz_i, \mRf_i, \dRfdz_i \right) \dz_i, \\
  A^{\ext\nuc} \left( \vec{x} \right) &= \sum_{i=0}^M L_{A^{\ext\nuc}}
  \left( \mz_i, \mRf_i, \dRfdz_i \right) \dz_i.
\end{align} \end{linenomath}
The integrands $L_{V^\nuc}$ and $L_{A^{\ext\nuc}}$ are given by
\eqref{eq:L_Vn} and \eqref{eq:L_Aen}, respectively, and
\begin{linenomath} \begin{align}
  \mz_i &= \frac{z_{i+1} + z_i}{2}, \\
  \mRf_i &= \frac{R^\film_{i+1} + R^\film_i}{2}, \\
  \dRfdz_i & = \frac{R^\film_{i+1} - R^\film_{i}}{\dz_i}, \label{eq:R_f_cd}\\
  \dz_i &= z_{i+1} - z_{i},
\end{align} \end{linenomath}
In addition, $R^\film_0 = R^\pore \pp{\posl}$, $R^\film_{M+1} =
R^\pore \pp{\posr}$, $z_0 = \posl$ and $z_{M+1} = \posr$.

The discretized Helmholtz energy for the film can now be calculated by
introducing into \eqref{eq:F_full_film_eliminated} the discrete
functionals $V^\nuc \left( \vec{x} \right)$ and $A^{\ext\nuc} \left(
\vec{x} \right)$, and the interfacial area $A^{\nuc\solid} \left(
\posl, \posr \right)$ as given by \eqref{eq:A_ns_film},
\begin{linenomath} \begin{align}
  F \left( \vec{y} \right) =\ & \nonumber \left\{\mu^\nuc - \mu^\ext
  \right\} N^\nuc - p^\ext V^\pore + \sigma^{\ext\solid} A^\pore +
  \mu^\ext N + \sigma^{\ext\nuc} A^{\ext\nuc} \left( \vec{x} \right)
  \\ & - \left\{ p^\nuc - p^\ext \right\} V^\nuc \left( \vec{x}
  \right) + \left\{ \sigma^{\nuc\solid} - \sigma^{\ext\solid} \right\}
  A^{\nuc\solid} \left( \posl, \posr \right).
\end{align} \end{linenomath}
Since, the quantities $V^\nuc$, $A^{\nuc\solid}$ and $A^{\nuc\ext}$
depend only on the variables contained in $\vec{x}$, and since
$V^\pore$, $A^\pore$ and $N$ are constants, the free variables of the
system are the elements of $\vec{x}$ and the number of particles in
the film $N^\nuc$,
\begin{linenomath} \begin{align}
  \vec{y} &= \nonumber \left[ y_1, y_2, ..., y_{M+3} \right]^\transp,
  \\ &= \left[ N^\nuc, \posl, \posr, R^\film_1, R^\film_2, ...,
    R^\film_M \right]^\transp.
\end{align} \end{linenomath}
The elements of the Jacobian vector can be obtained from
\eqref{eq:F_full_film_eliminated_diff} as
\begin{linenomath} \begin{align}
  \pd{F}{y_1} &= \mu^\nuc - \mu^\ext,
\end{align} \end{linenomath}
and, for $i \in \left[ 2,...,M+3\right]$,
\begin{linenomath} \begin{align}
  \pd{F}{y_i} =& -\left( p^\nuc - p^\ext \right) \pd{V^\nuc}{y_i} +
  \left( \sigma^{\nuc\solid} - \sigma^{\ext\solid} \right)
  \pd{A^{\nuc\solid}}{y_i} + \sigma^{\nuc\ext} \pd{A^{\nuc\ext}}{y_i}.
\end{align} \end{linenomath}
The Hessian matrix can be found by further differentiation. The
procedure used for calculating the derivatives of $V^\nuc$ and
$A^{\nuc\ext}$ w.r.t.\ $y_i$ is documented in the supplementary
material. By a derivation similar to that above, we obtain the grand
potential $\Omega$, its Jacobian vector and its Hessian matrix.

The strategy to solve the discrete problem is to use the stationary
state obtained from solving the ODE as described in
Section~\ref{sec:numfilm_ode} as initial guess. The stationary state
for the discrete problem is then found by solving for the vector
$\vec{y}^*$ for which the discrete Jacobian vector is zero,
\begin{linenomath} \begin{align}
  \label{eq:dFdy_0_film}
  \left.\od{F}{\vec{y}}\right|_{\vec{y}^*} &= \vec{0}.
\end{align} \end{linenomath}
Since we have an expression for the Hessian matrix of $F$ and a very
good initial guess for the solution to \eqref{eq:dFdy_0_film}, this
non-linear system of equations is solved with Newton's method with few
number of iterations. A convergence study is reported in
\ref{app:film_convergence}, which shows that the film profiles obtained
by the discrete method converge to those obtained by solving the
Euler--Lagrange equation as the grid size $M$ is increased. For a
thorough exposition of the discrete approach, we refer to the
supplementary material.

\begin{figure*}[th]
  \centering
  \begin{subfigure}[b]{0.46\textwidth}
    \centering
    \includegraphics[width=\textwidth]{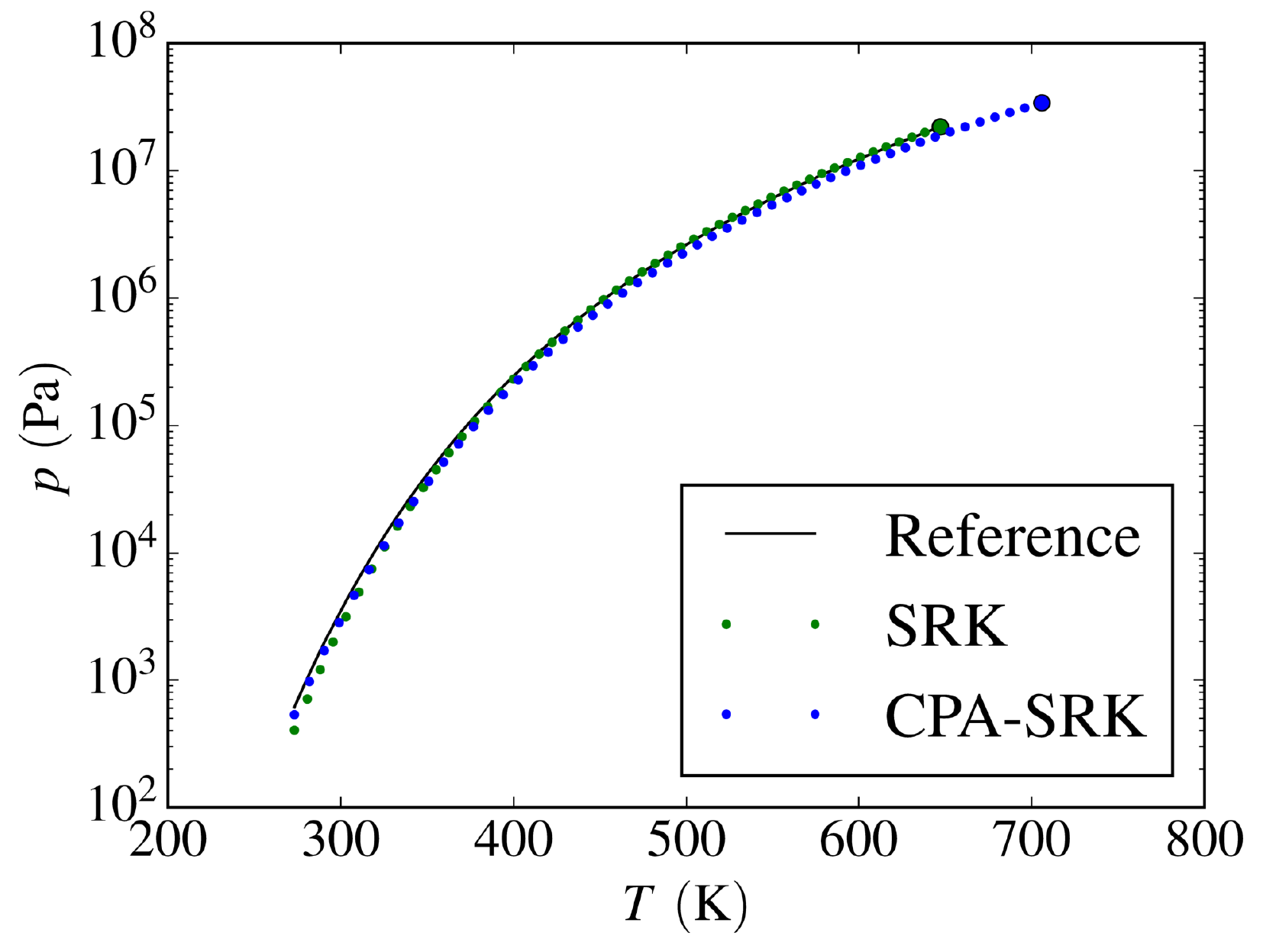}
    \caption{}
  \end{subfigure}
  \begin{subfigure}[b]{0.47\textwidth}
    \centering
    \includegraphics[width=\textwidth]{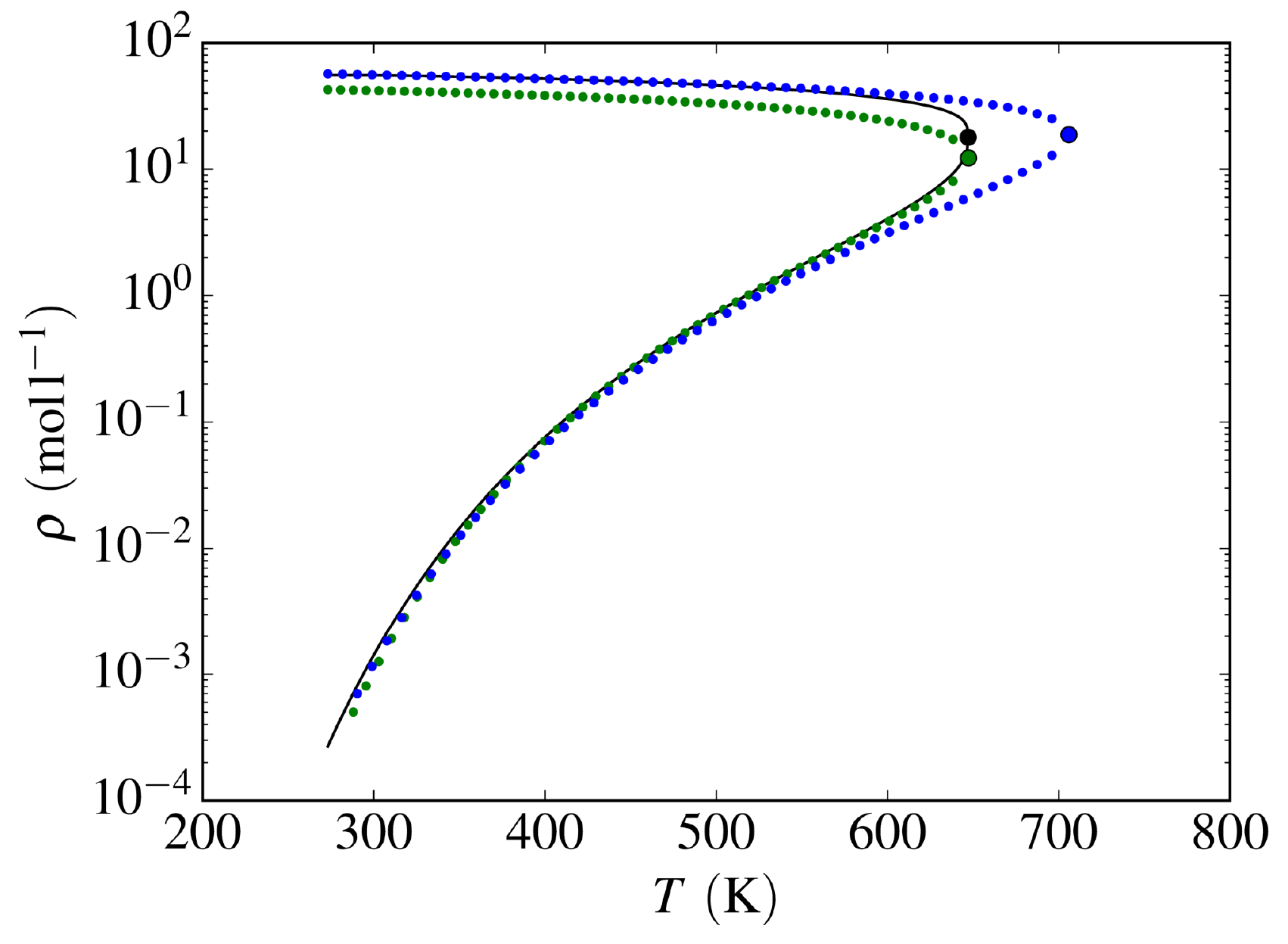}
    \caption{}
  \end{subfigure}
  \caption{Saturation properties of water, as predicted by SRK (green)
    and CPA-SRK (blue). Reference data from
    \cite{NISTChemistryWebBook} are shown for comparison
    (black). Compared to SRK, CPA-SRK is more inaccurate in the
    critical region, but has superior predictions of liquid densities
    in the temperatures around \SI{300}{\kelvin}.}
  \label{fig:H2O_saturation}
\end{figure*}

\begin{figure}[th]
  \centering
    \includegraphics[width=0.48\textwidth]{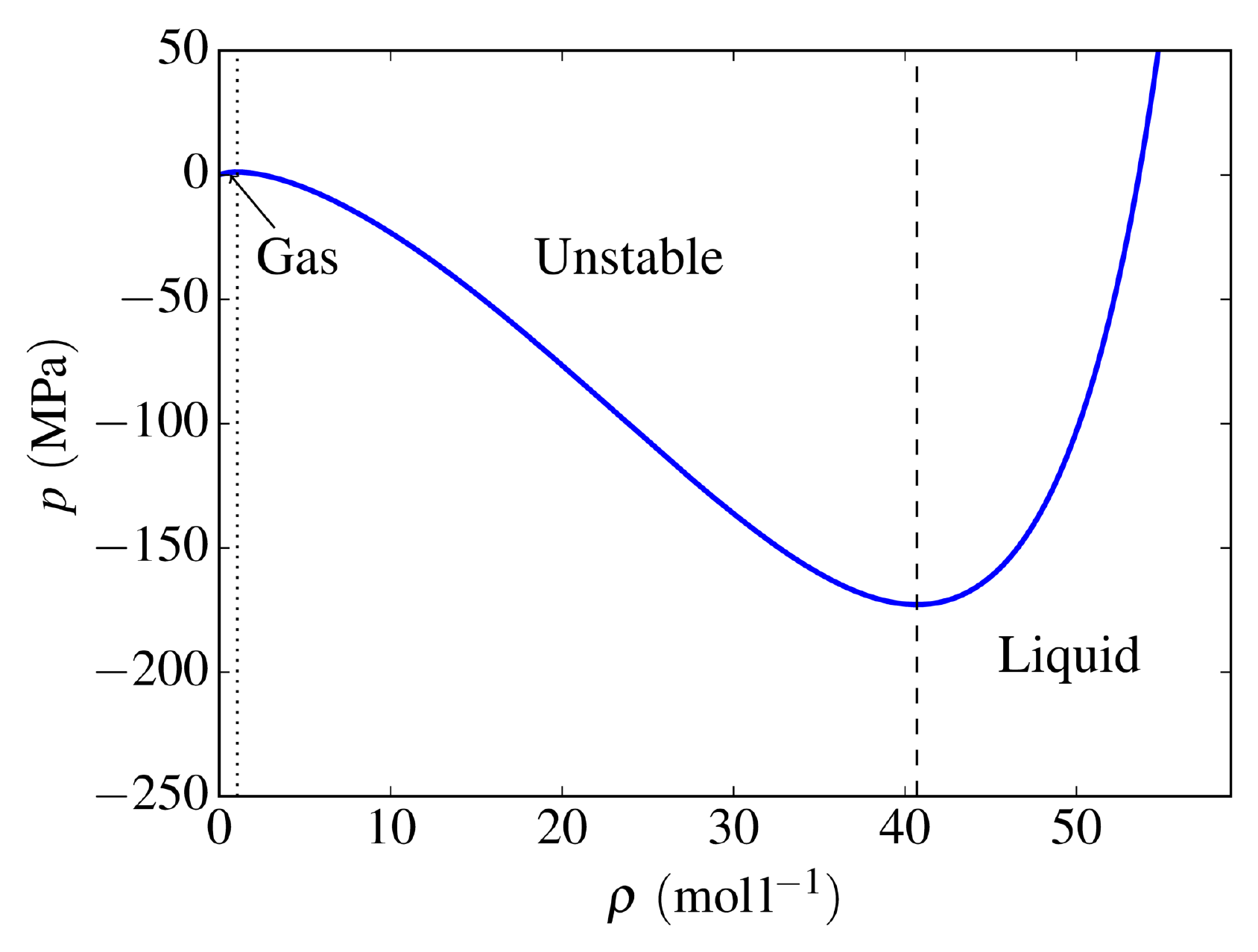}
    \caption{CPA-SRK isotherm (solid blue) for water at
      $\SI{358}{\kelvin}$. The densities at the gas spinodal
      (dotted line) and liquid spinodal (dashed line) are also
      indicated. Phases with densities between the two spinodals are
      unstable.}
    \label{fig:H2O_isotherm}
\end{figure}

\subsection{Stability analysis}
Any stationary state identified using the procedures described above,
and defined by the vector $\vec{y}^*$, will be stationary for both $F$
and $\Omega$ and, therefore, in both a closed and an open pore. Since
the Jacobian vector is $\vec{0}$, the change in, say, $F$ in response
to a small perturbation $\dd \vec{y}$ of $\vec{y}^*$ can be expressed
as
\begin{linenomath} \begin{align}
  \dd F &= \dd \vec{y}^\transp
  \left.\odd{F}{\vec{y}}\right|_{\vec{y}^*} \dd \vec{y},
\end{align} \end{linenomath}
where $\dd^2 F / \dd \vec{y}^2$ is the Hessian matrix. For the film,
we calculate the Hessian matrices using the discrete description, see
Section~\ref{sec:film_discr}. The symmetric Hessian matrix can be
decomposed into
\begin{linenomath} \begin{align}
  \left.\odd{F}{\vec{y}}\right|_{\vec{y}^*} &= \vec{Q} \vec{\Lambda}
  \vec{Q}^{\transp},
\end{align} \end{linenomath}
where $\vec{\Lambda}$ is the diagonal matrix eigenvalues and $\vec{Q}$
is a matrix where column $i$ is the eigenvector $\vec{q}_i$ (with unit
length in the $L_2$-norm) associated with eigenvalue $\lambda_i$. The
eigenvectors are orthogonal, since the Hessian is symmetric.

A stationary state $\vec{y^*}$ corresponds to a minimum in $F$ and is
considered locally stable in the closed pore if all eigenvalues of the
Hessian are positive. If one or more of the eigenvalues are negative,
$\dd \vec{y}$ can be taken in direction of the corresponding
eigenvectors $\vec{q}_i$ (or $-\vec{q}_i$) to give a negative $\dd
F$. The stationary state is thus not a minimum in the Helmholtz energy
and it is therefore unstable. Analogous considerations apply for
$\Omega$ and stability in open pores.

The eigenvectors that correspond to the negative eigenvalues give
information about the direction of the perturbations that lead to a
reduction in the energy and make the system unstable. For both the
adsorbed droplet/bubble and the films, we observe (see
Section~\ref{sec:results}) two distinct classes of instabilities that
we name (1) translation and (2) condensation/evaporation. Translation
instabilities are perturbations where the $\nuc$-phase moves along the
$z$-axis and only a small number of particles are transferred to/from
the $\ext$-phase(s). 
For condensation/evaporation instabilities on the other hand, the
$\nuc$-phase expands or contracts while exchanging particles with the
$\ext$-phase(s), without shifting its center of mass. 

Eigenvectors and eigenvalues were calculated using \verb|eigh| from
\verb|numpy|'s \verb|linalg| module \cite{SciPy}. This function uses
the \verb|*syevd| routines from LAPACK, which compute the eigenvalues
and eigenvectors of symmetric matrices \cite{lapack}.

\section{Results}
\label{sec:results}
In the following, we will discuss the thermodynamic stability of the
heterogeneous structures illustrated in
Figure~\ref{fig:fluid_configurations}. The focus will be on the
influence of pore size, the fluid-solid interaction, as captured by a
finite contact angle $\alpha$, and the difference in thermodynamic
stability between closed and open systems. We restrict the discussion
to two pore sizes, $L^\pore=\SI{10}{\micro\meter}$ and $L^\pore =
\SI{0.01}{\micro\meter}$. Despite the small size of both pores, we
will refer to the \SI{10}{\micro\meter}-pore as large and the
\SI{0.01}{\micro\meter}-pore as small.

Water at $\SI{358}{\kelvin}$ will be used as example, inspired by the
operational conditions of a proton-exchange membrane fuel
cell\footnote{Condensation of liquid water in such fuel cells may
  block reactant flow paths and is severely detrimental to their
  performance. It is therefore of interest to know if, say, a
  path-blocking adsorbed droplet or a liquid film is the equilibrium
  configuration under the chosen operating
  conditions.}~\cite{Bednarek2017}. The thermodynamic properties of
water are described by the cubic-plus-association modification to the
Soave--Redlich--Kwong EOS (CPA-SRK). In
Figure~\ref{fig:H2O_saturation}, the saturation properties of water as
described by both the Soave--Redlich--Kwong EOS (SRK) and CPA-SRK are
plotted together with reference data
from~\cite{NISTChemistryWebBook}. Compared to SRK, CPA-SRK is more
inaccurate in the critical region, but has superior prediction of
liquid densities at lower temperatures and is therefore the preferred
choice here.

The CPA-SRK isotherm for water at $\SI{358}{\kelvin}$ is shown in
Figure~\ref{fig:H2O_isotherm}, where the gas and liquid spinodals are
indicated by vertical lines. A homogeneous phase with a density
between the two spinodals is thermodynamically unstable. The isotherm
shows that the EOS predicts stable or metastable stretched liquid
phases down to pressures of \SI{-173}{\mega\pascal}, consistent with
the findings of \citet{Caupin_2012} and \citet{Azouzi_2013}.

In the models presented in Section~\ref{sec:models}, the energies of
the gas-liquid interface $\sigma^{\gas\liquid}$, the gas-solid
interface $\sigma^{\gas\solid}$ and the liquid-solid interface
$\sigma^{\liquid\solid}$ are necessary input parameters. The
gas-liquid surface tension of water at $\SI{358}{\kelvin}$ is
$\SI{0.0616}{\newton\per\meter}$~\cite{NISTChemistryWebBook}. With
this value in place, it is only the difference between the gas-solid
and gas-liquid interfacial tensions that is of physical significance
in the models. We therefore set $\sigma^{\liquid\solid} = 0$ and
subsequently use $\sigma^{\gas\liquid}$ and the specified contact
angle $\alpha$ to calculate $\sigma^{\gas\solid}$ by use of Young's
equation \eqref{eq:yl}.

\subsection{Pore with a free droplet or bubble}
\begin{figure*}[htbp]
  \centering
  \begin{subfigure}[b]{0.48\textwidth}
    \centering
    \includegraphics[width=\textwidth]{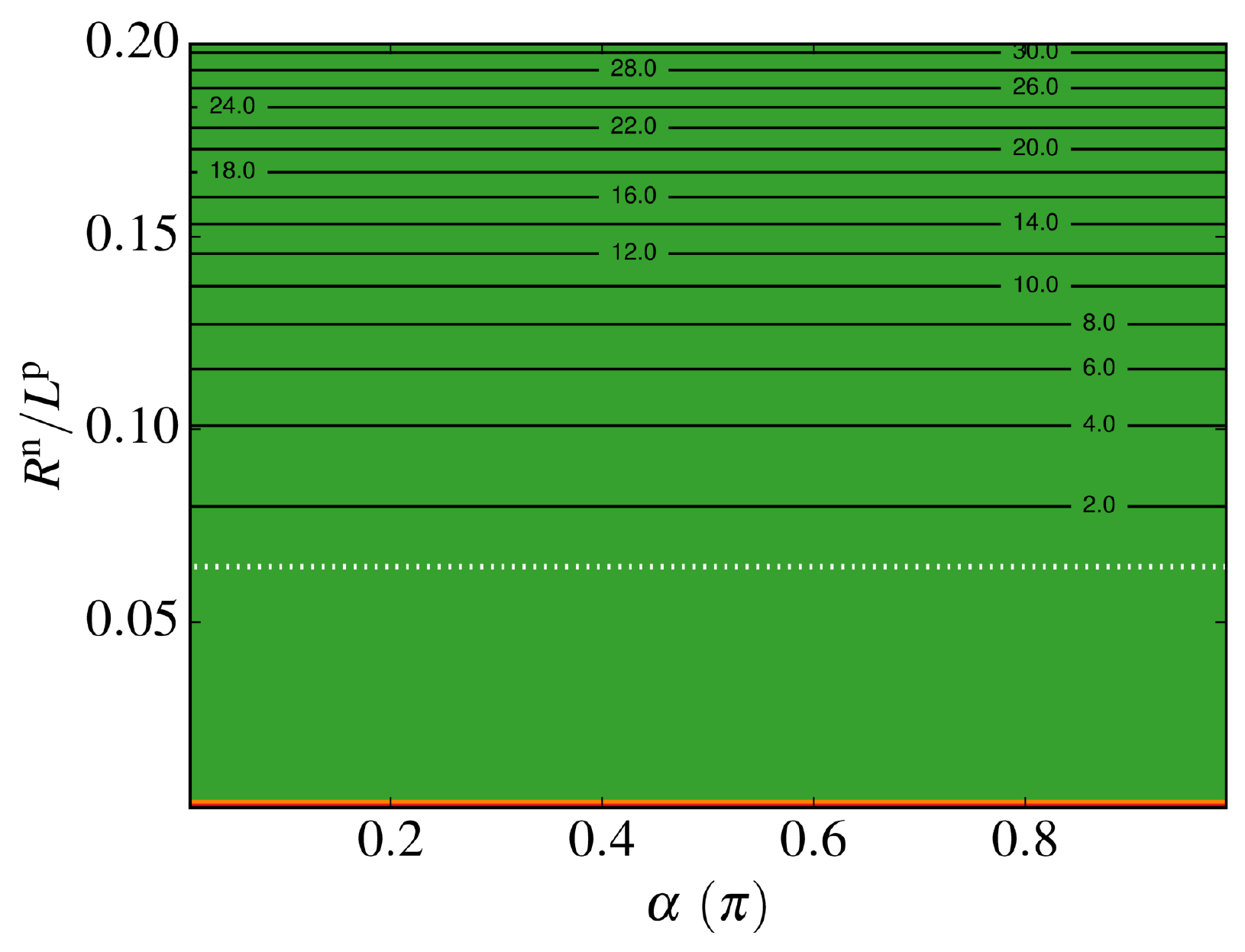}
    \caption{Free droplets in closed pore of \SI{10}{\micro\meter}}
    \label{fig:free_droplet_closed_10um}
  \end{subfigure}~
  \begin{subfigure}[b]{0.48\textwidth}
    \centering
    \includegraphics[width=\textwidth]{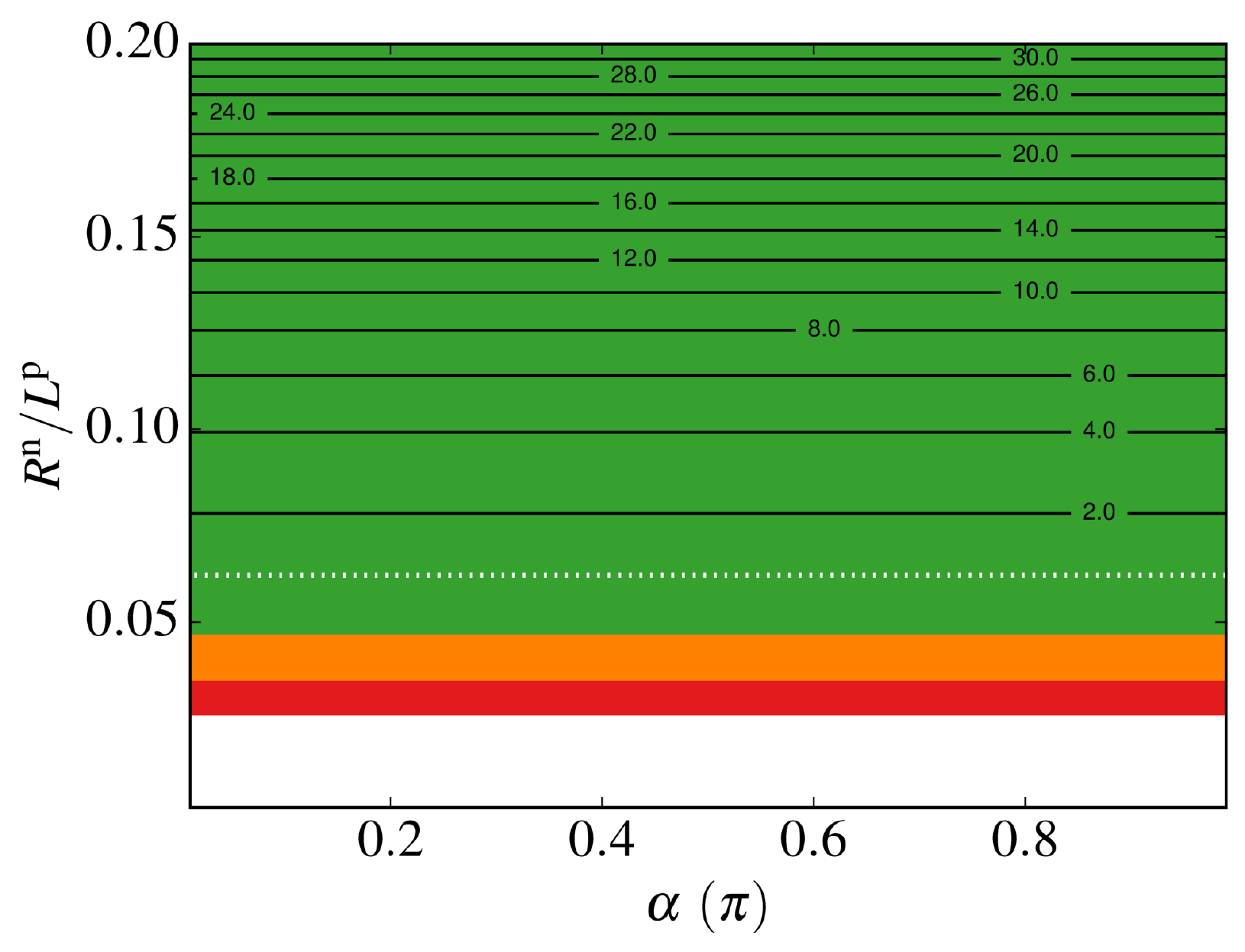}
    \caption{Free droplets in closed pore of \SI{0.01}{\micro\meter}}
    \label{fig:free_droplet_closed_0.01um}
  \end{subfigure}
  \begin{subfigure}[b]{0.48\textwidth}
    \centering
    \includegraphics[width=\textwidth]{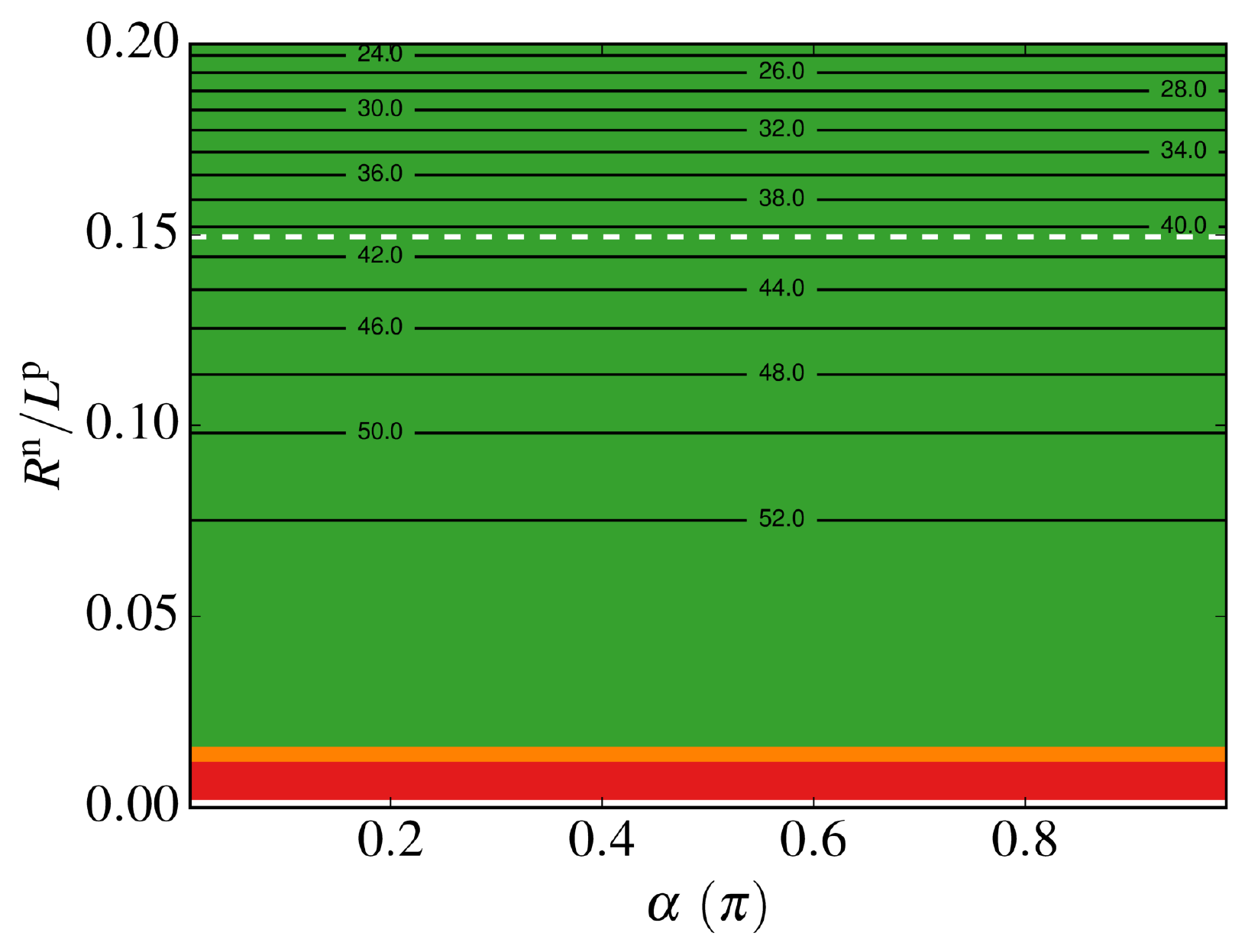}
    \caption{Free bubbles in closed pore of \SI{10}{\micro\meter}}
    \label{fig:free_bubble_closed_10um}
  \end{subfigure}~
  \begin{subfigure}[b]{0.48\textwidth}
    \centering
    \includegraphics[width=\textwidth]{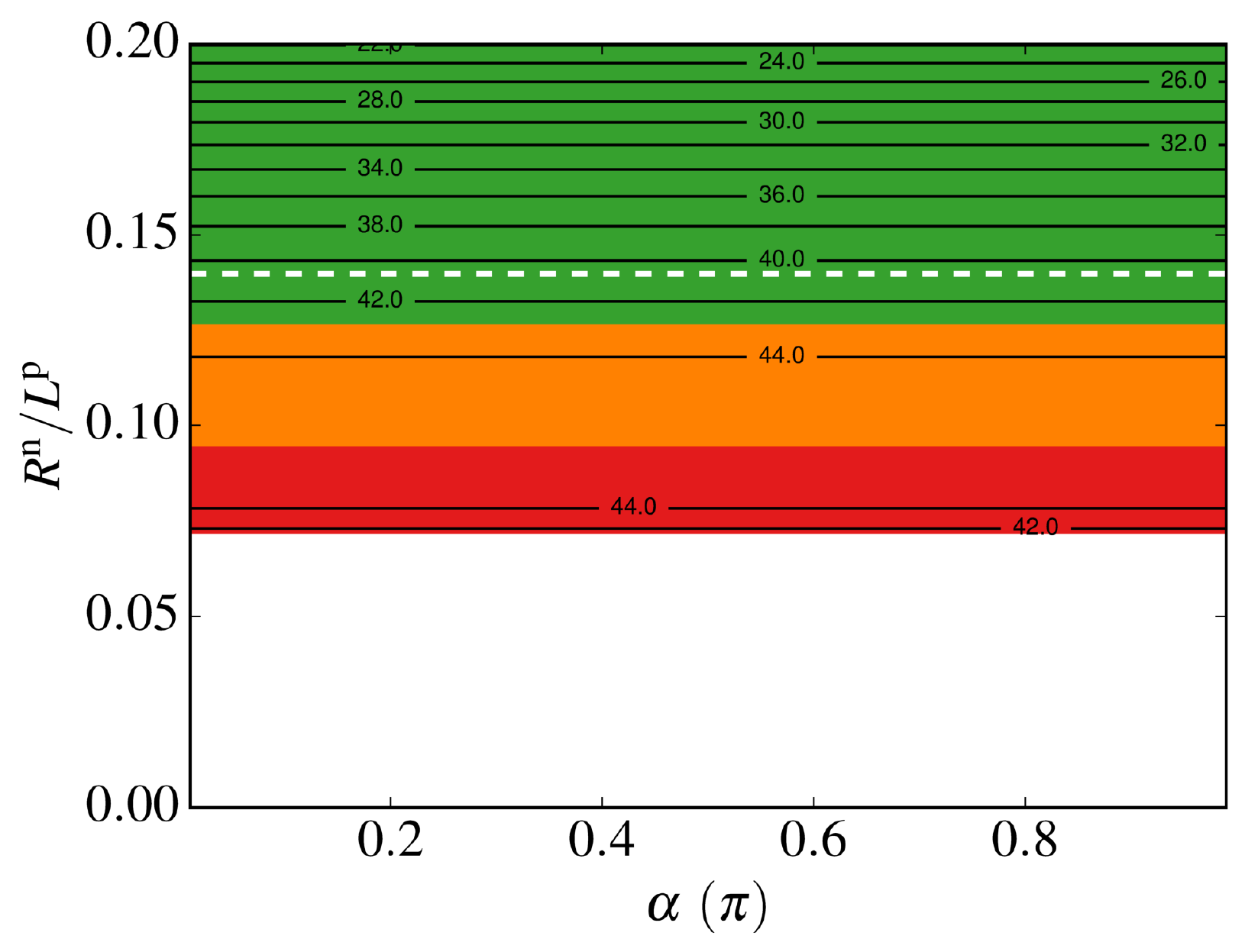}
    \caption{Free bubbles in closed pore of \SI{0.01}{\micro\meter}}
    \label{fig:free_bubble_closed_0.01um}
  \end{subfigure}
  \caption{Stability maps for (a), (b) free droplets and (c), (d) free
    bubbles in closed pores of lengths (a), (c) \SI{10}{\micro\meter}
    and (b), (d) \SI{0.01}{\micro\meter}. The black contour lines
    indicate the total fluid density in \si{\mol\per\litre}. Unstable
    configurations are red, stable configurations are green and
    configurations that are locally stable, but have a larger
    Helmholtz energy than if the pore were filled with a single
    homogeneous phase, are orange. The gas spinodal density of
    \SI{1.05}{\mol\per\litre} is drawn as a dotted white line and the
    liquid spinodal of \SI{40.7}{\mol\per\litre} is shown as a dashed
    white line. No homogeneous phase filling the entire pore can exist
    for densities between the spinodals.}
  \label{fig:free_droplet_closed}
\end{figure*}

The thermodynamic stability of free droplets and bubbles in a closed
pore has been studied in previous
works~\cite{Yang1985,Wilhelmsen2014super,Wilhelmsen2015cav}. Our
results for water at \SI{358}{\kelvin} are shown in
Figure~\ref{fig:free_droplet_closed}. The stability is here mapped out
in terms of the relative droplet/bubble size $R^\nuc/L^\pore$ and
contact angle $\alpha$. The largest bubble radius that has been
considered equals the radius of the pore at the widest point.

As expected, the thermodynamic stability of free bubbles and droplets
is independent of contact angle. For both pore sizes, large bubbles
and droplets are stable and have lower Helmholtz energies than if the
pores were filled with a homogeneous phase (with the same number of
particles). As $R^\nuc$ decreases however, the configurations first
become metastable with respect to the homogeneous phase and,
eventually, unstable. The reason is that, as the volume of the
$\nuc$-phase becomes smaller, the reduction in energy from having both
a liquid and a gas phase does not compensate for the energy cost of
the gas-liquid interface. These findings are consistent with those
of~\citet{Wilhelmsen_2014}. They considered droplets and bubbles in a
spherical container, where the interfacial energy between the
container and the $\ext$-phase was zero. The non-zero energy of the
$\ext\solid$-interface adds a constant term to the Helmholtz energy
that does not change the local stability w.r.t.\ to the analysis
performed by~\citet{Wilhelmsen_2014}. This is also the reason why the
contact angle does not affect stability. However, it is necessary to
include the contribution from the fluid-solid interface when comparing
the energy of the free droplet/bubble configurations with other
configurations such as adsorbed droplet/bubbles and films.

The thermodynamic stability of free droplets and bubbles in a closed
system changes with pore size, which can be seen by comparing
Figure~\ref{fig:free_droplet_closed_10um} with
\ref{fig:free_droplet_closed_0.01um} for the droplets and
Figure~\ref{fig:free_bubble_closed_10um} with
\ref{fig:free_bubble_closed_0.01um} for the bubbles. In the white
region in the bottom part of the figures, the droplet and bubble
radius becomes so small that the pressure difference needed to satisfy
the Young--Laplace relation is too large to conform with equal
chemical potential between the phases. The limiting factor for the
droplets is the gas spinodal and for the bubbles it is the liquid
spinodal. The density range where the bubbles and droplets are
thermodynamically stable (the green regions) or metastable (orange
regions) decreases with pore size. This is due to the
superstabilization of the homogeneous phase that occurs in small
pores. We refer to \cite{Wilhelmsen2014super} for an elaborate
discussion of this topic.

The white dashed and dotted lines in
Figure~\ref{fig:free_droplet_closed} mark the states where the total
density in the pore reaches the spinodal limits, and the homogeneous
fluid becomes unstable. Beyond this limit, the fluid will
spontaneously decompose into two phases.

In agreement with previous work, we find that free droplets and
bubbles are unstable in the open systems~\cite{Yang1985}, also in
the presence of a solid-fluid interface.

\begin{figure*}[htbp]
  \centering
  \begin{subfigure}[b]{0.48\textwidth}
    \centering
    \includegraphics[width=\textwidth]{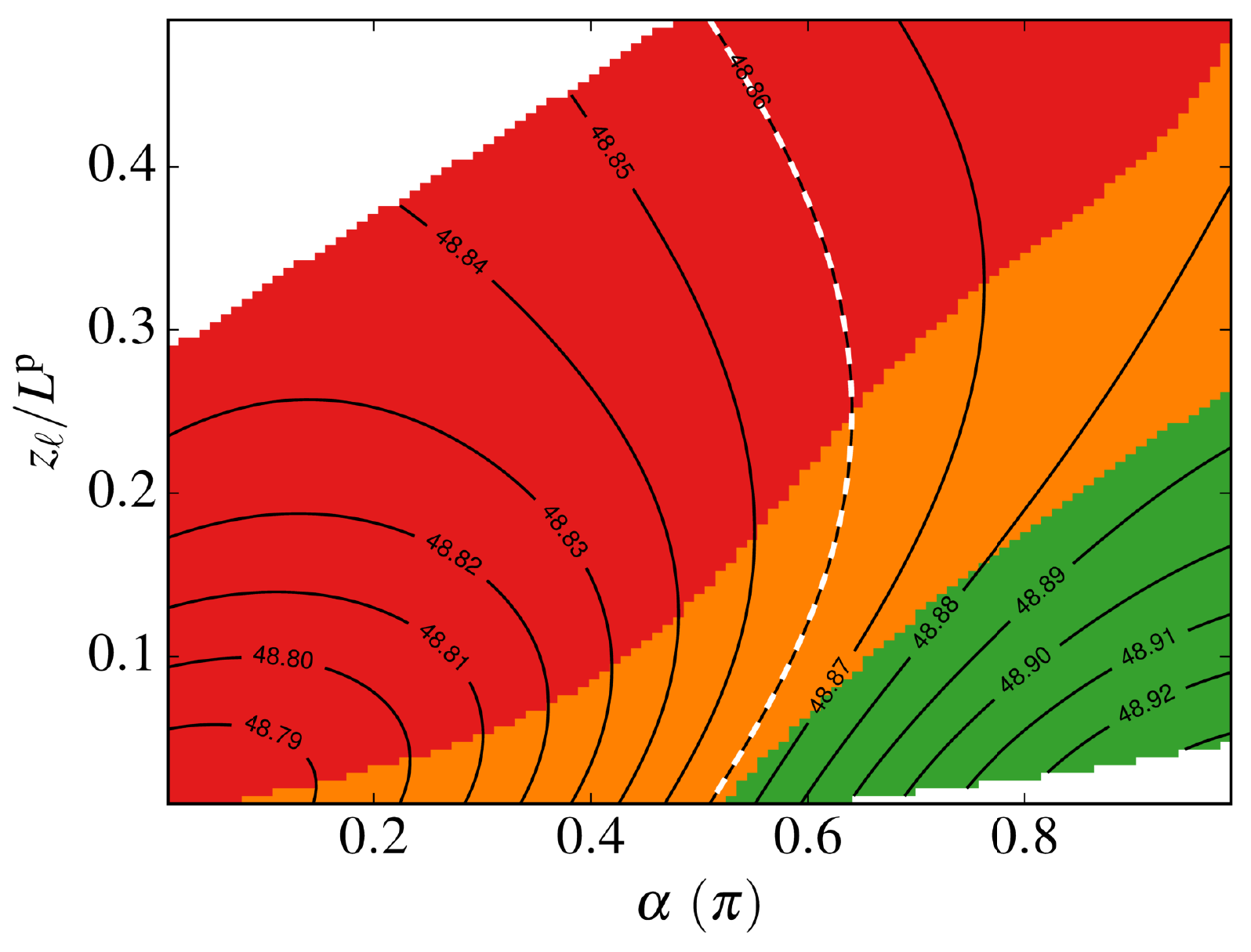}
    \caption{Adsorbed droplet in open pore of \SI{10}{\micro\meter}}
    \label{fig:adsorbed_droplet_open_10um}
  \end{subfigure}~
    \begin{subfigure}[b]{0.48\textwidth}
    \centering
    \includegraphics[width=\textwidth]{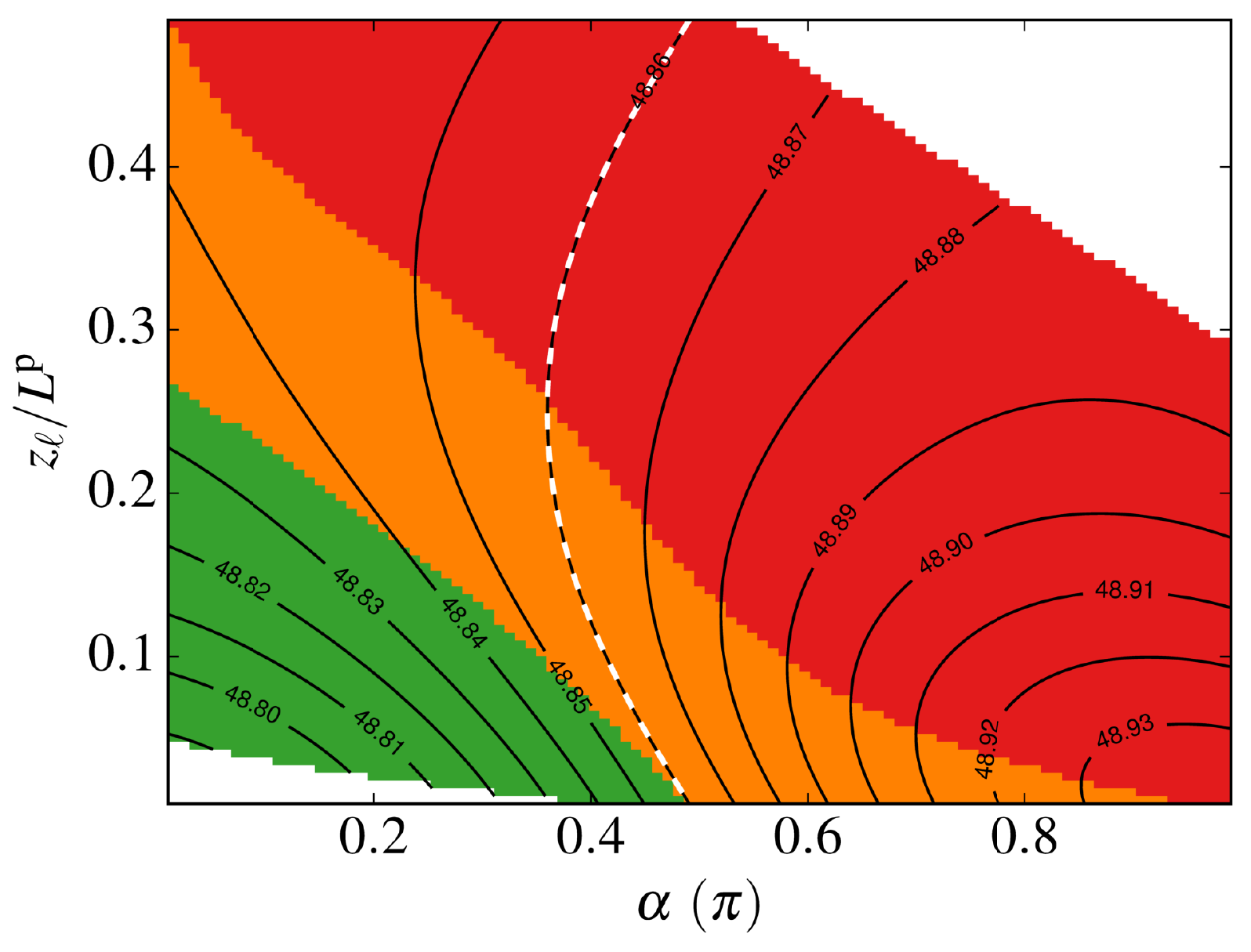}
    \caption{Adsorbed bubbles in open pore of \SI{10}{\micro\meter}}
    \label{fig:adsorbed_bubble_open_10um}
  \end{subfigure}
  \begin{subfigure}[b]{0.48\textwidth}
    \centering
    \includegraphics[width=\textwidth]{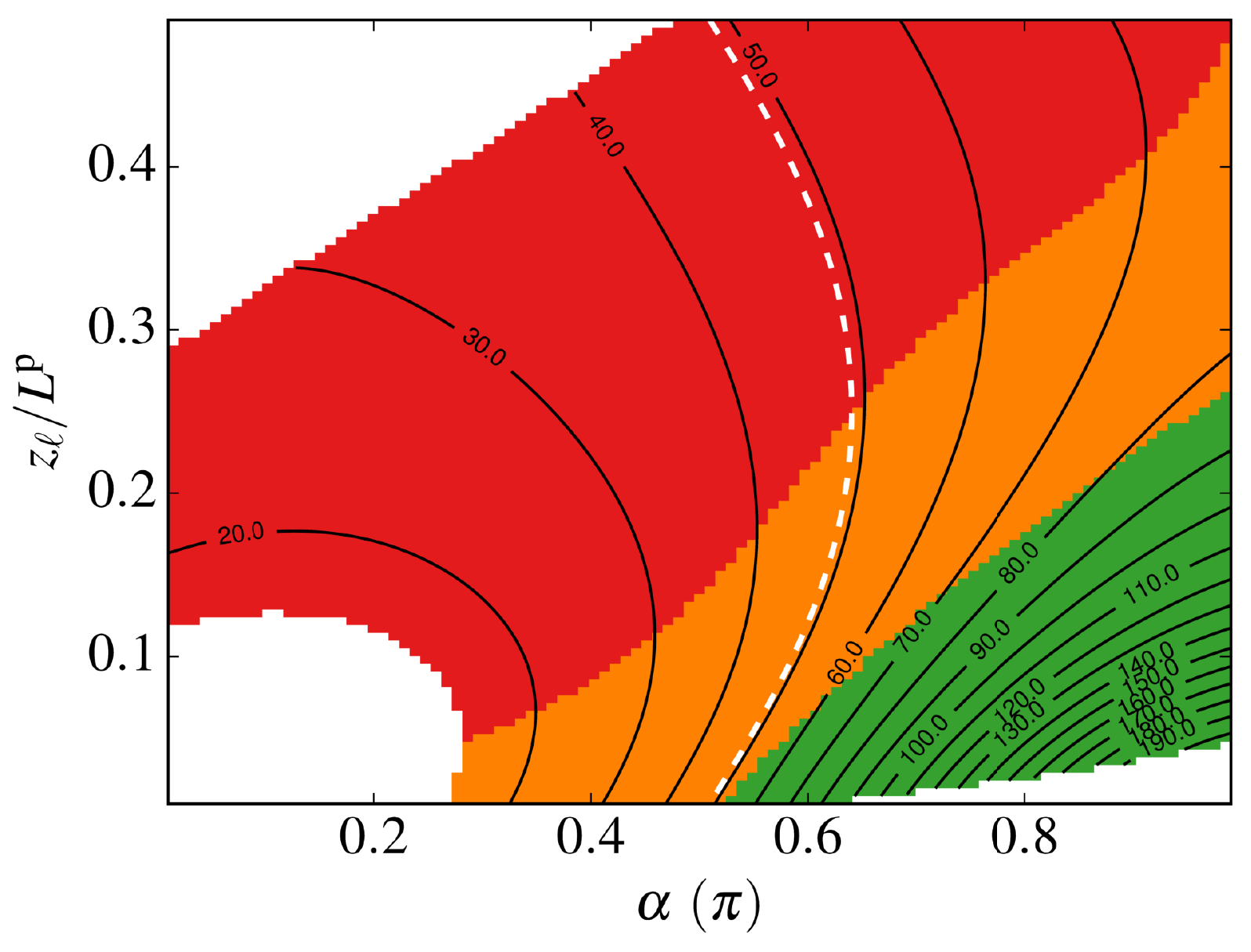}
    \caption{Adsorbed droplets in open pore of \SI{0.01}{\micro\meter}}
    \label{fig:adsorbed_droplet_open_10nm}
  \end{subfigure}~
  \begin{subfigure}[b]{0.48\textwidth}
    \centering
    \includegraphics[width=\textwidth]{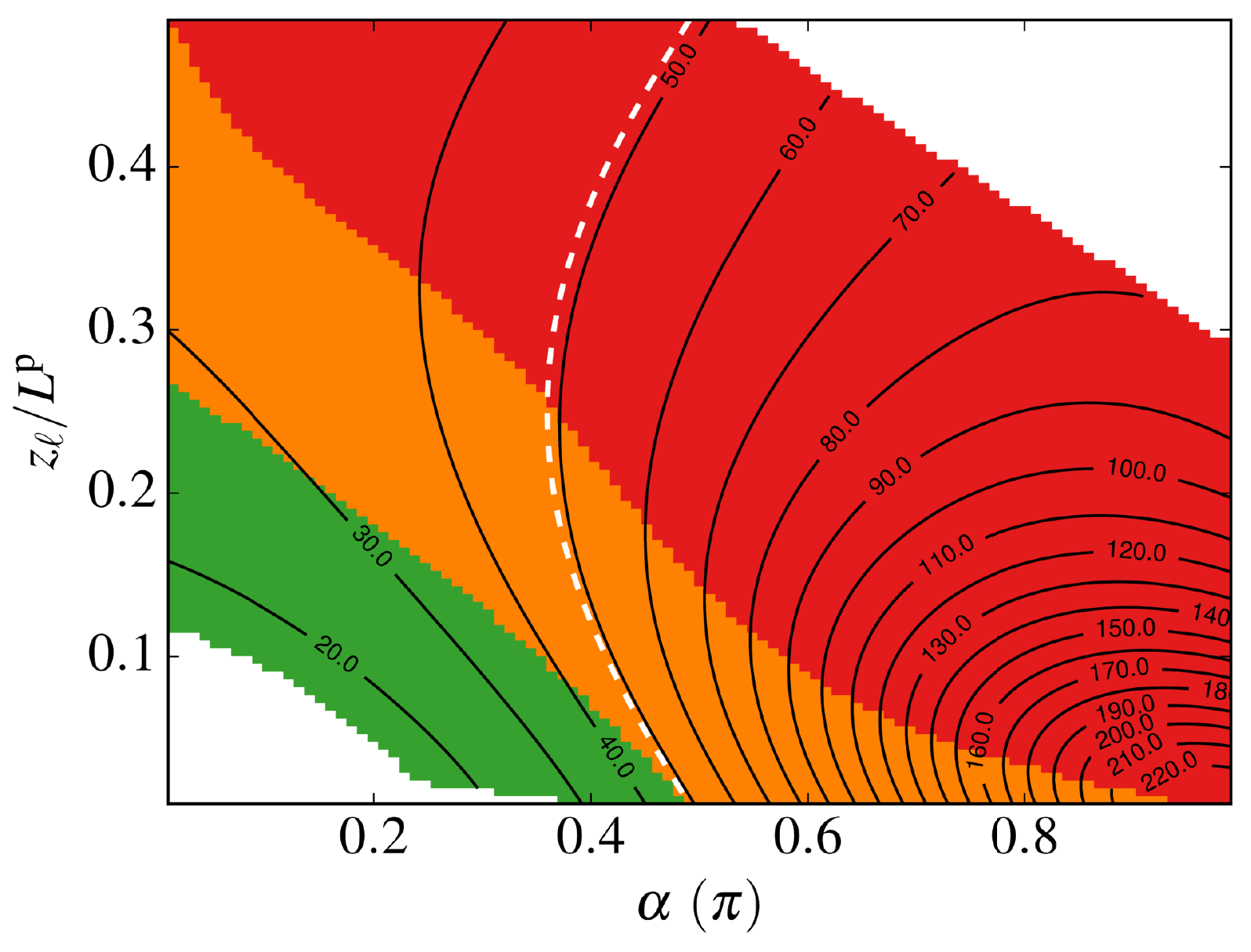}
    \caption{Adsorbed bubbles in open pore of \SI{0.01}{\micro\meter}}
    \label{fig:adsorbed_bubble_open_10nm}
  \end{subfigure}
  \begin{subfigure}[b]{0.48\textwidth}
    \centering
    \includegraphics[width=\textwidth]{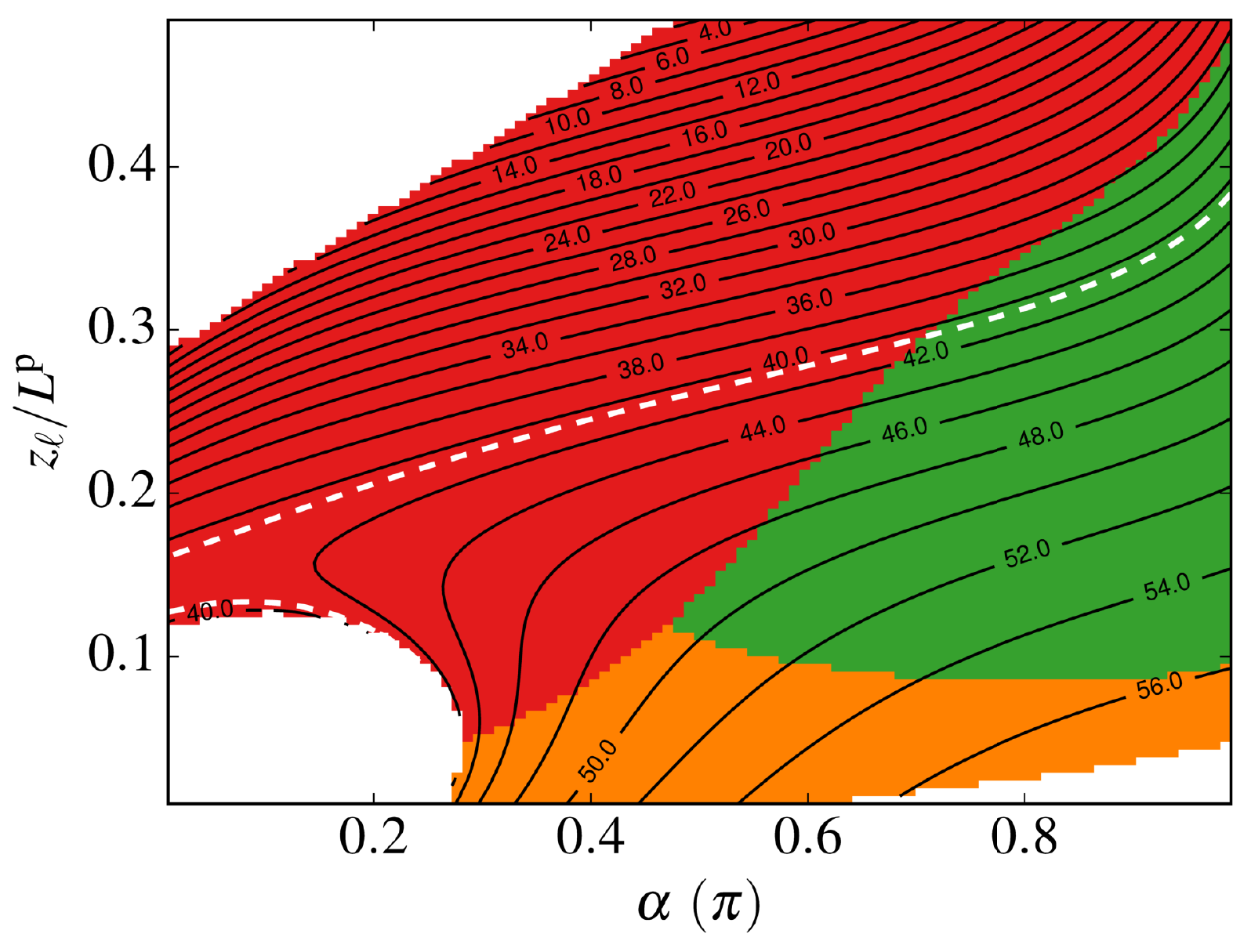}
    \caption{Adsorbed droplets in closed pore of \SI{0.01}{\micro\meter}}
    \label{fig:adsorbed_droplet_closed_10nm}
  \end{subfigure}~
  \begin{subfigure}[b]{0.48\textwidth}
    \centering
    \includegraphics[width=\textwidth]{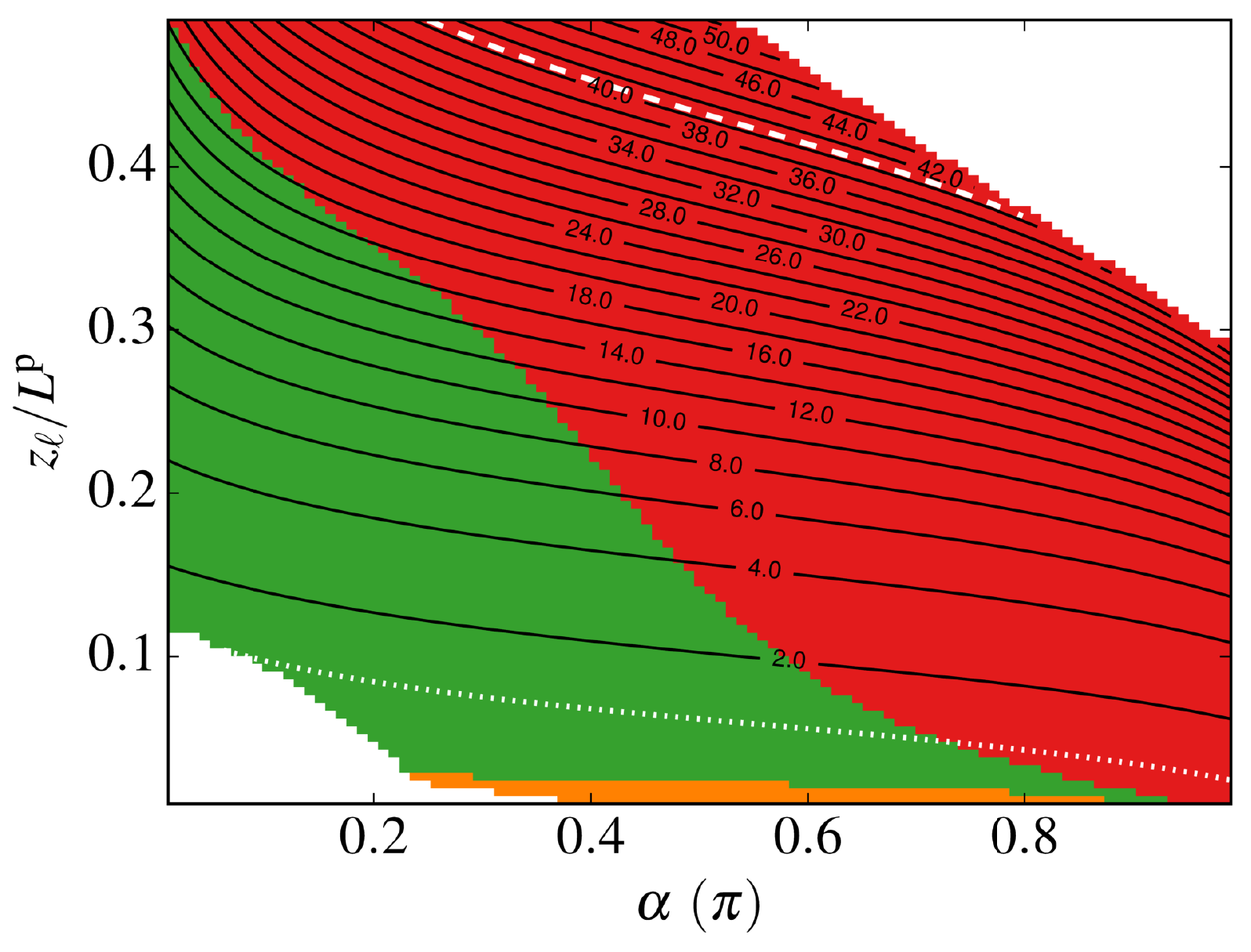}
    \caption{Adsorbed bubbles in closed pore of \SI{0.01}{\micro\meter}}
    \label{}
  \end{subfigure}
  \caption{Stability maps of (left column) adsorbed droplets and
    (right column) adsorbed bubbles in open and closed pores with 
    lengths \SI{10}{\micro\meter} and \SI{0.01}{\micro\meter}. For
    the closed pores, the black contour lines indicate the total fluid
    density in \si{\mol\per\litre}. Further, the gas spinodal density
    of \SI{1.05}{\mol\per\litre} is drawn as a dotted white line and
    the liquid spinodal of \SI{40.7}{\mol\per\litre} is shown as a
    dashed white line. No homogeneous phase filling the entire pore
    can exist for densities between the spinodals. For the open pores,
    the black contour lines indicate the gas phase pressure in
    \si{\kilo\pascal}. Also, the bulk saturation pressure of
    \SI{48.86}{\kilo\pascal} is shown as a dashed white line. For all
    maps, unstable configurations are red, stable configurations are
    green and configurations that are locally stable, but have a
    larger energy than if the pore were filled with a single
    homogeneous phase, are orange.}
  \label{fig:adsorbed_droplet_bubble}
\end{figure*}

\subsection{Pore with an adsorbed droplet or bubble}
\label{sec:results_adsorbed_bubbledroplet}
When the solid-fluid interfacial energy is lower than the gas-liquid
interfacial tension, the bubbles and droplets can lower their energies
by adsorbing to the pore walls, as shown in
Figure~\ref{fig:pore_with_adsorbed_droplet}. The thermodynamic
stability of adsorbed bubbles and droplets is mapped out in
Figure~\ref{fig:adsorbed_droplet_bubble} in terms of the position of
the left meniscus $\posl$ and the contact angle $\alpha$ for open and
closed pores with lengths \SI{10}{\micro\meter} and
\SI{0.01}{\micro\meter}.

In contrast to the free droplets and bubbles, the absorbed droplets
and bubbles can be stable both in open and closed pores. The general
trend is that adsorbed droplets are thermodynamically stable if the
liquid contact angle is high (non-wetting), while bubbles are stable
when the liquid contact angle is low (wetting), and the range of
stability depends on the value of the contact angle. We find that this
behavior depends strongly on the pore geometry. An in-depth discussion
of the influence of pore morphology on the thermodynamic stability
however, is beyond the scope of the present work.

We have analyzed in detail the regions where the adsorbed droplets and
bubbles become unstable. The eigenvectors associated with the negative
Hessian eigenvalues in these regions reveal that an instability
w.r.t.\ to translation of the $\nuc$-phase along the $z$-axis, i.e.\ a
mechanical instability, is present in the unstable regions of all the
adsorbed bubble and droplet configurations. A perturbation of the
position of the droplet/bubble leads to a net force that moves it
further in the direction of the perturbation, not back to the original
position as for the stable droplets/bubbles. The Hessian matrices of
the open systems have an additional negative eigenvalue in the
unstable regions. The second negative eigenvalue is associated with
condensation/evaporation. This instability is also present for the
droplet in the small closed pore, when the liquid pressures become
large and negative. We have included figures that display where these
instabilities appear in the supplementary material. Comparing the
stability regions of the droplets (left column in
Figure~\ref{fig:adsorbed_droplet_bubble}) with those of the bubbles
(right column Figure~\ref{fig:adsorbed_droplet_bubble}), these are
clearly anti-symmetric. This is because the droplets with contact
angles $\alpha$ are mechanically identical to bubbles with contact
angle $\pi - \alpha$.

The white regions in the stability maps for the
\SI{10}{\micro\meter}-pore represent configurations where no
stationary state can be found because the two menisci would have
intersected or extended outside the pore. Such configurations are
unfeasible and are not considered in the analysis. The same is true
for the small pores. In addition, the large curvatures of some menisci
in the small pores result in large, negative liquid
pressures. Figure~\ref{fig:H2O_isotherm} shows that the liquid
spinodal poses a lower limit to how large the negative pressures of
the liquid phase can be. Configurations with a lower pressure than the
liquid spinodal are unfeasible.


Figure~\ref{fig:adsorbed_droplet_closed_10nm} shows results for a
closed pore of the small kind containing an adsorbed droplet. It has a
considerable region where the droplets are metastable w.r.t.\ a pore
with a homogeneous liquid phase (with the same number of
particles). In a large closed pore, the thermodynamic stability map
looks the same, except that the metastable region is stable. A figure
can be found in the supplementary material. The reason for the
appearance of the metastable region in the small pore is as for the
free droplet/bubble; when the volumes of the bulk phases become
smaller, the reduction in energy from having both a gas and a liquid
does not compensate for the energy associated with the gas-liquid
interface. The adsorbed bubble configurations display a similar
behavior, where large bubbles (relative to pore size) become unstable
w.r.t.\ a homogeneous gas phase. A crucial difference from the free
bubbles/droplets, however, is that the transition from stable to
metastable depends on the contact angle.

The orange regions for the open pores indicate where the
adsorbed/droplet bubble configurations are locally stable, but have a
larger grand canonical energy than if the pore were filled with a
homogeneous phase with the same intensive properties as the
$\ext$-phase. The location of these regions does not change much when
pore size is reduced, which is evident by comparing
Figures~\ref{fig:adsorbed_bubble_open_10um} to
\ref{fig:adsorbed_bubble_open_10nm} for the bubbles, and
Figures~\ref{fig:adsorbed_droplet_open_10um} to
\ref{fig:adsorbed_droplet_open_10nm} for the droplets.

Comparing the adsorbed bubbles in the large open pore
(Figure~\ref{fig:adsorbed_bubble_open_10um}) with those in the small
pore (Figure~\ref{fig:adsorbed_bubble_open_10nm}), we observe that the
variation of gas pressure is much larger in the small pores. In
particular when the liquid phase is sufficiently wetting, the gas
pressure is lower than the bulk saturation pressure, and much more so
in the small pore. Although the external phase is liquid in
Figure~\ref{fig:adsorbed_bubble_open_10um} and
Figure~\ref{fig:adsorbed_bubble_open_10nm}, these observation are
closely related to the phenomenon of capillary condensation in open
pores with an external gas phase. When the liquid phase is wetting, a
lower water vapor pressure is required for water to condense in the
pore than in the bulk. A much lower vapor pressure is required in the
small pores, meaning that water will preferentially condense in small
over large pores. A study of capillary condensation and the effects
pore size and shape can be done with the models and methods presented
here, which is a possible topic for future work.

\subsection{Pore with a thick films of liquid or gas}
\begin{figure*}[htbp]
  \centering
  \begin{subfigure}[b]{0.48\textwidth}
    \centering
    \includegraphics[width=\textwidth]{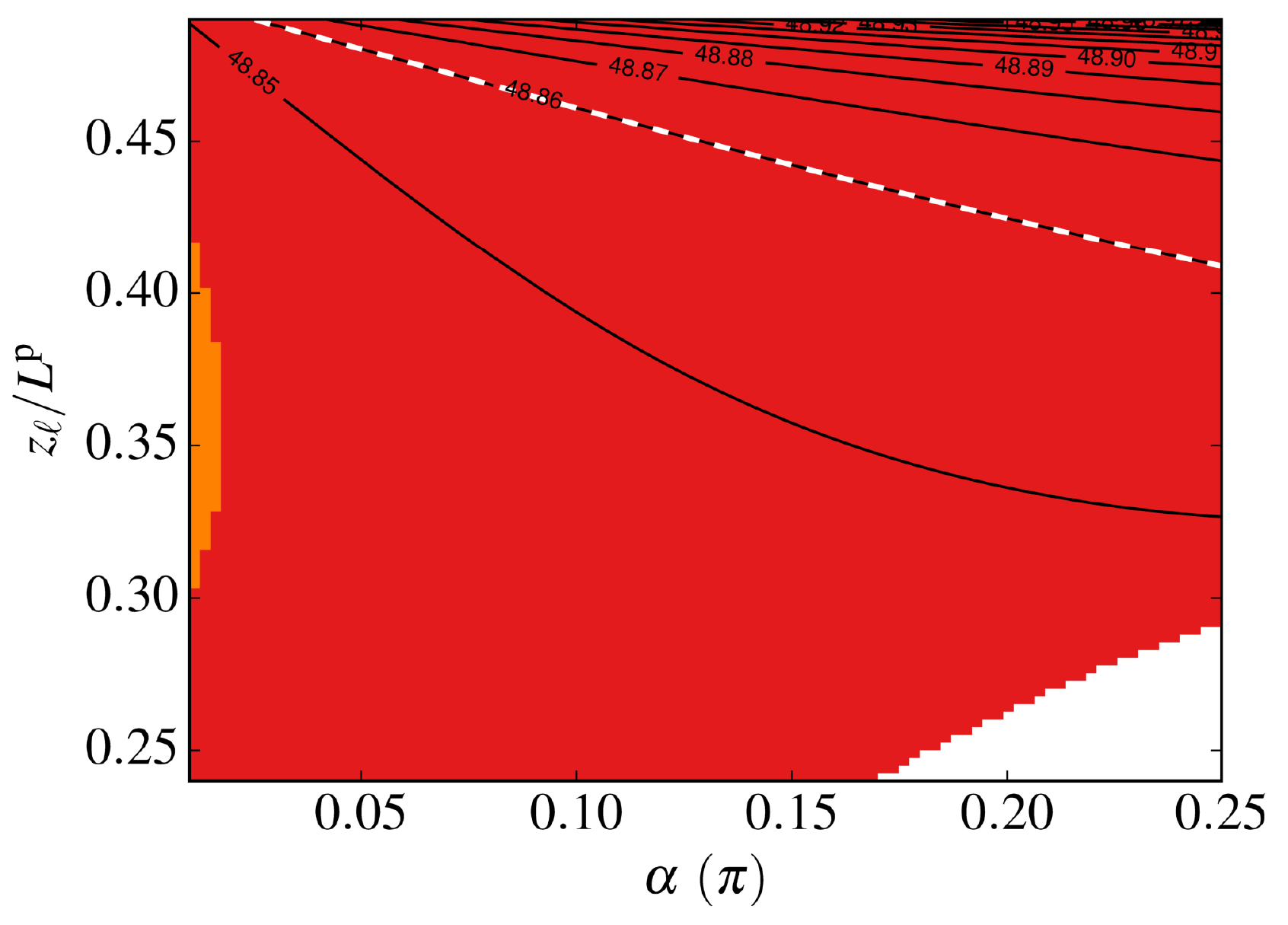}
    \caption{Liquid films in open pore of \SI{10}{\micro\meter}}
    \label{fig:liquid_film_open_10um}
  \end{subfigure}~  
  \begin{subfigure}[b]{0.48\textwidth}
    \centering
    \includegraphics[width=\textwidth]{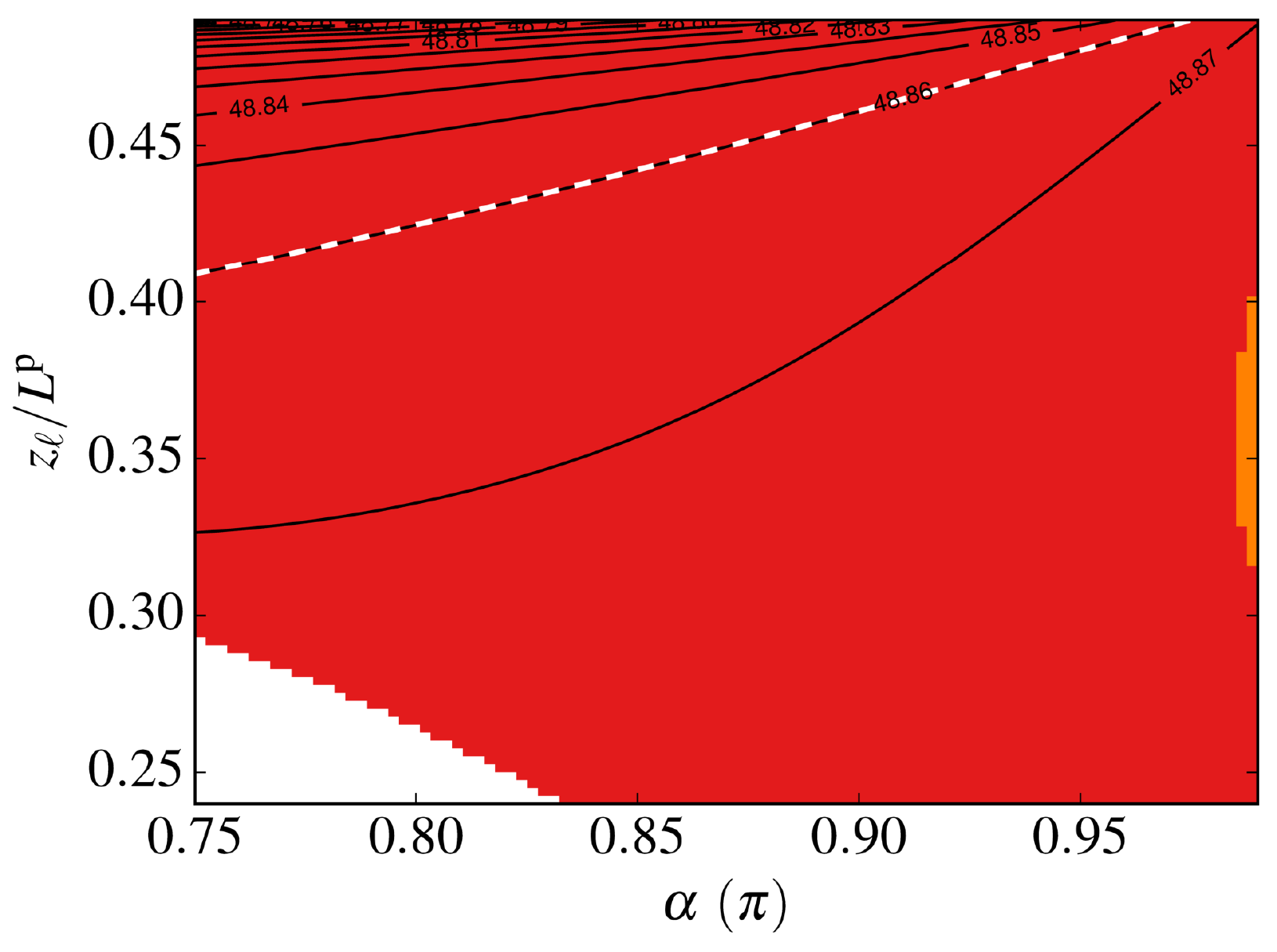}
    \caption{Gas films in open pore of \SI{10}{\micro\meter}}
    \label{fig:gas_film_open_10um}
  \end{subfigure}
  \begin{subfigure}[b]{0.48\textwidth}
    \centering
    \includegraphics[width=\textwidth]{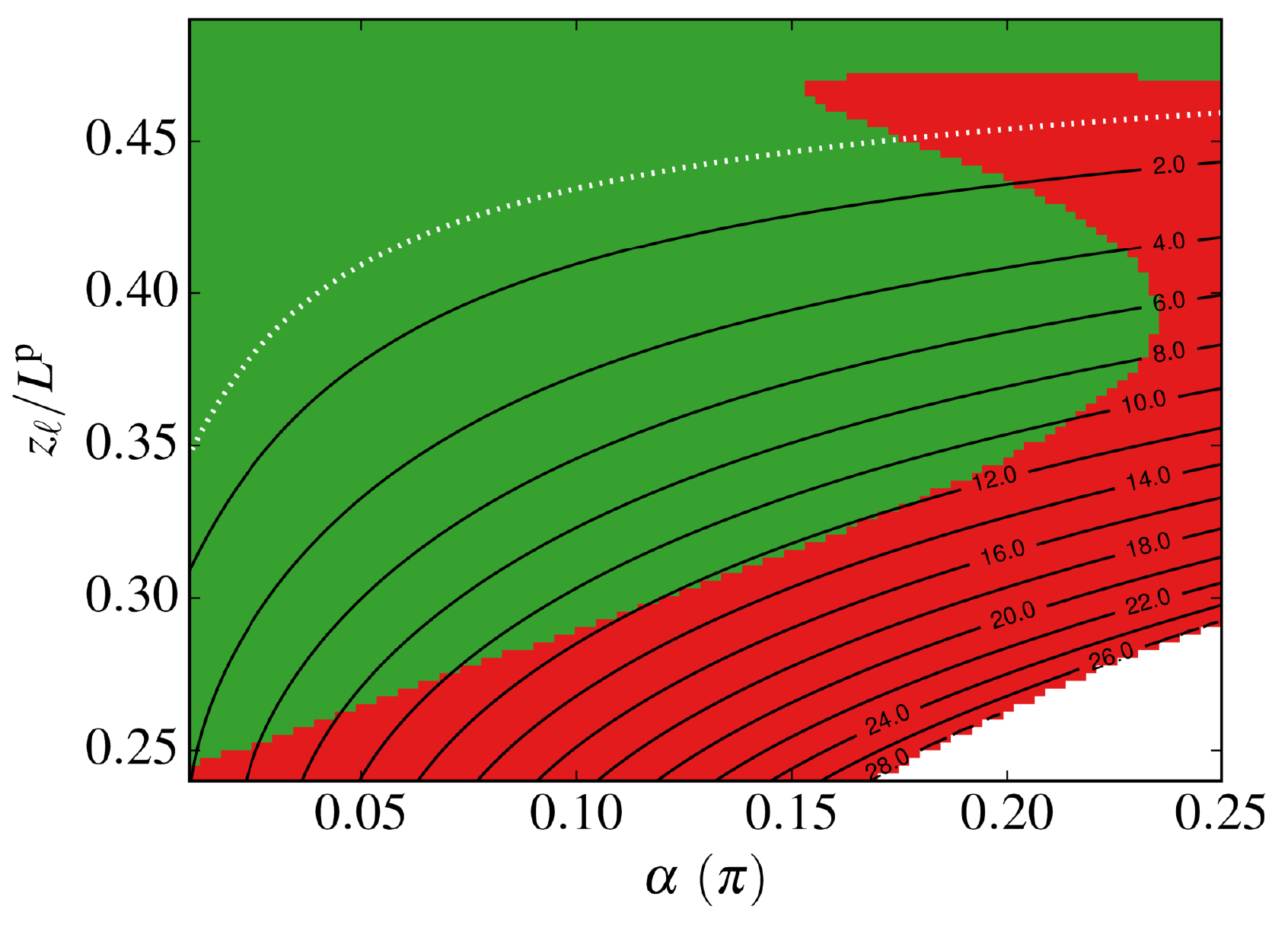}
    \caption{Liquid films in closed pore of \SI{10}{\micro\meter}}
    \label{fig:liquid_film_closed_10um}
  \end{subfigure}~  
  \begin{subfigure}[b]{0.48\textwidth}
    \centering
    \includegraphics[width=\textwidth]{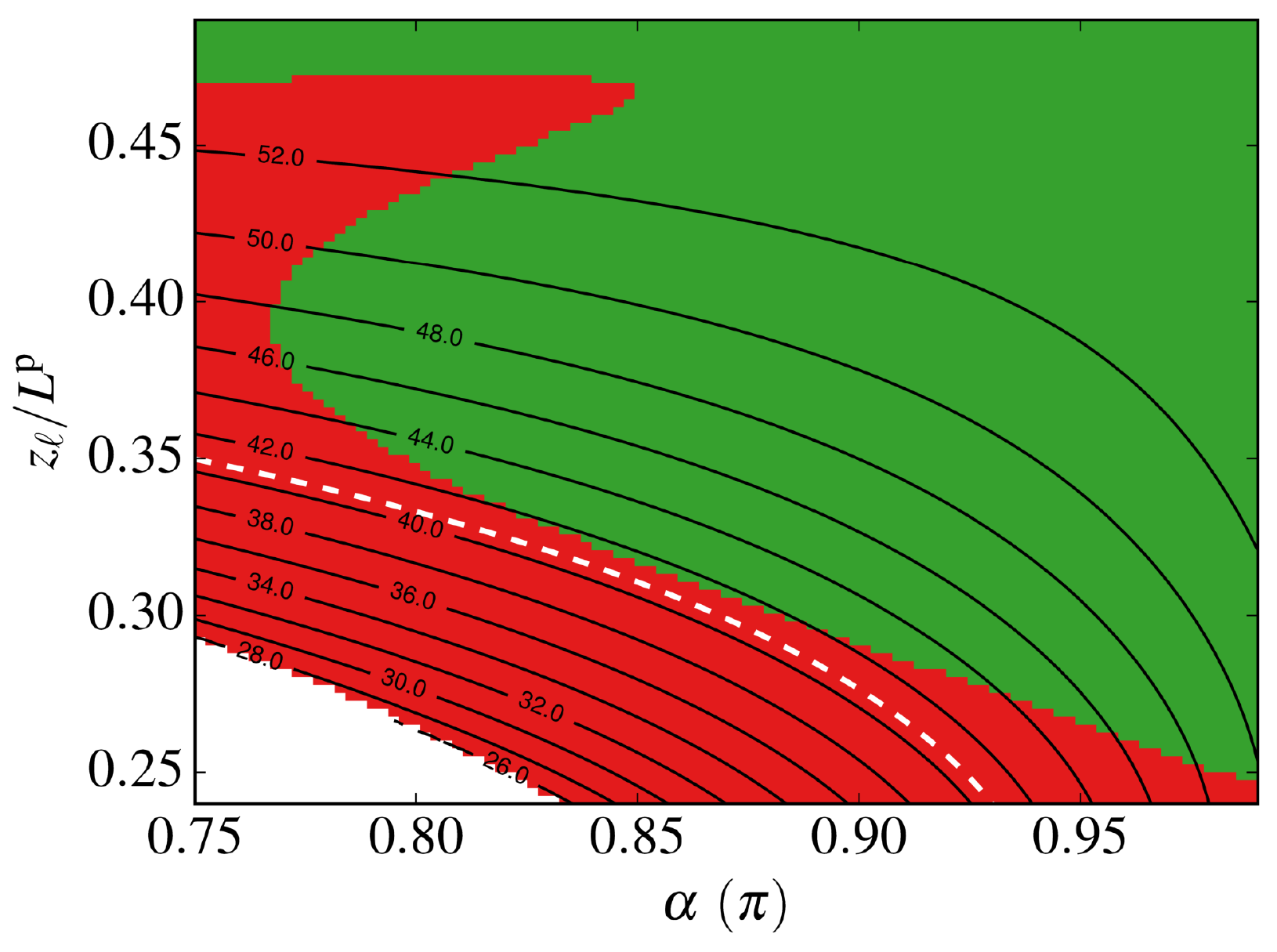}
    \caption{Gas films in closed pore of \SI{10}{\micro\meter}}
    \label{fig:gas_film_closed_10um}
  \end{subfigure}
    \begin{subfigure}[b]{0.48\textwidth}
    \centering
    \includegraphics[width=\textwidth]{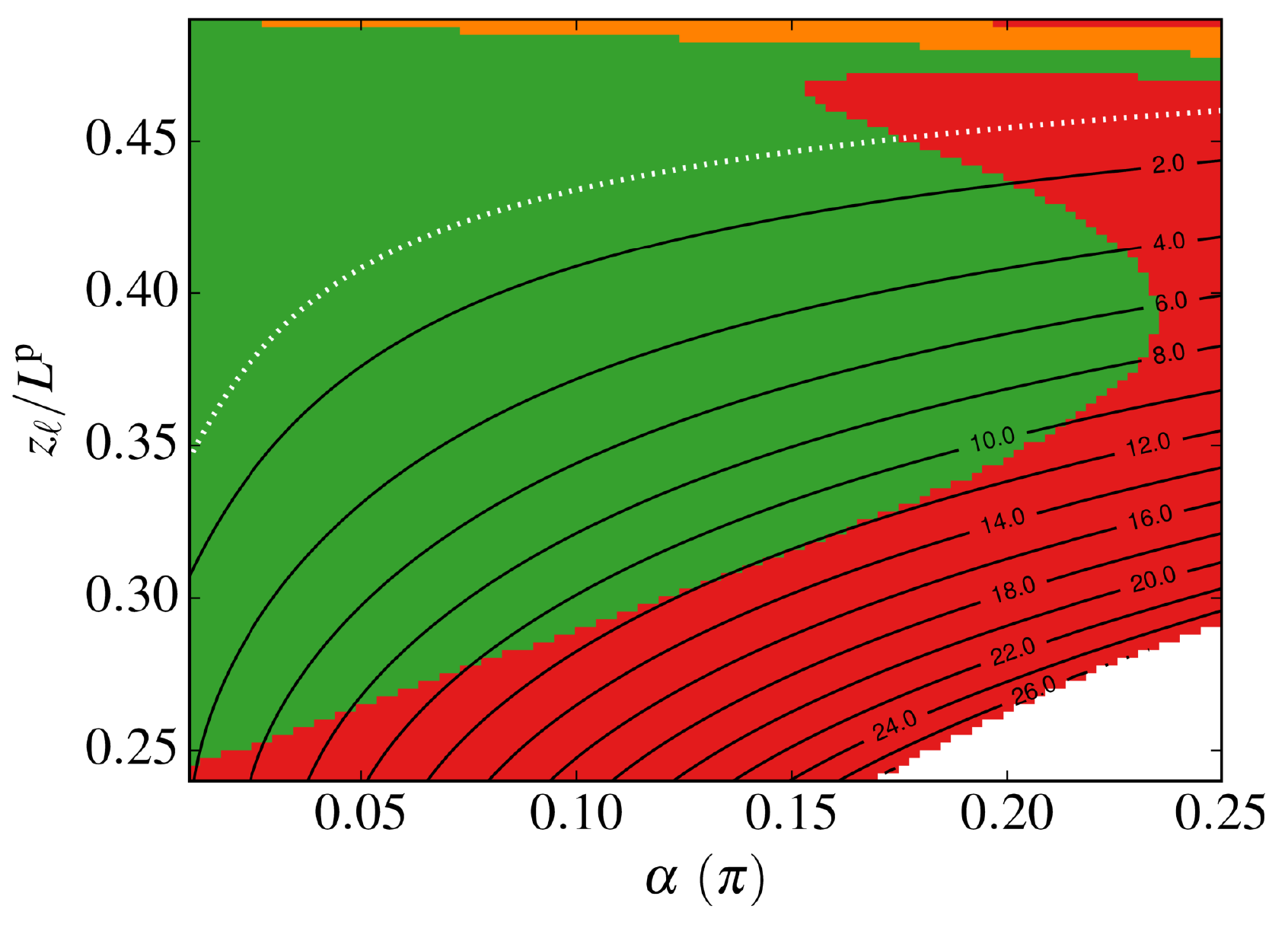}
    \caption{Liquid films in closed pore of \SI{0.01}{\micro\meter}}
    \label{fig:liquid_film_closed_10nm}
  \end{subfigure}~  
  \begin{subfigure}[b]{0.48\textwidth}
    \centering
    \includegraphics[width=\textwidth]{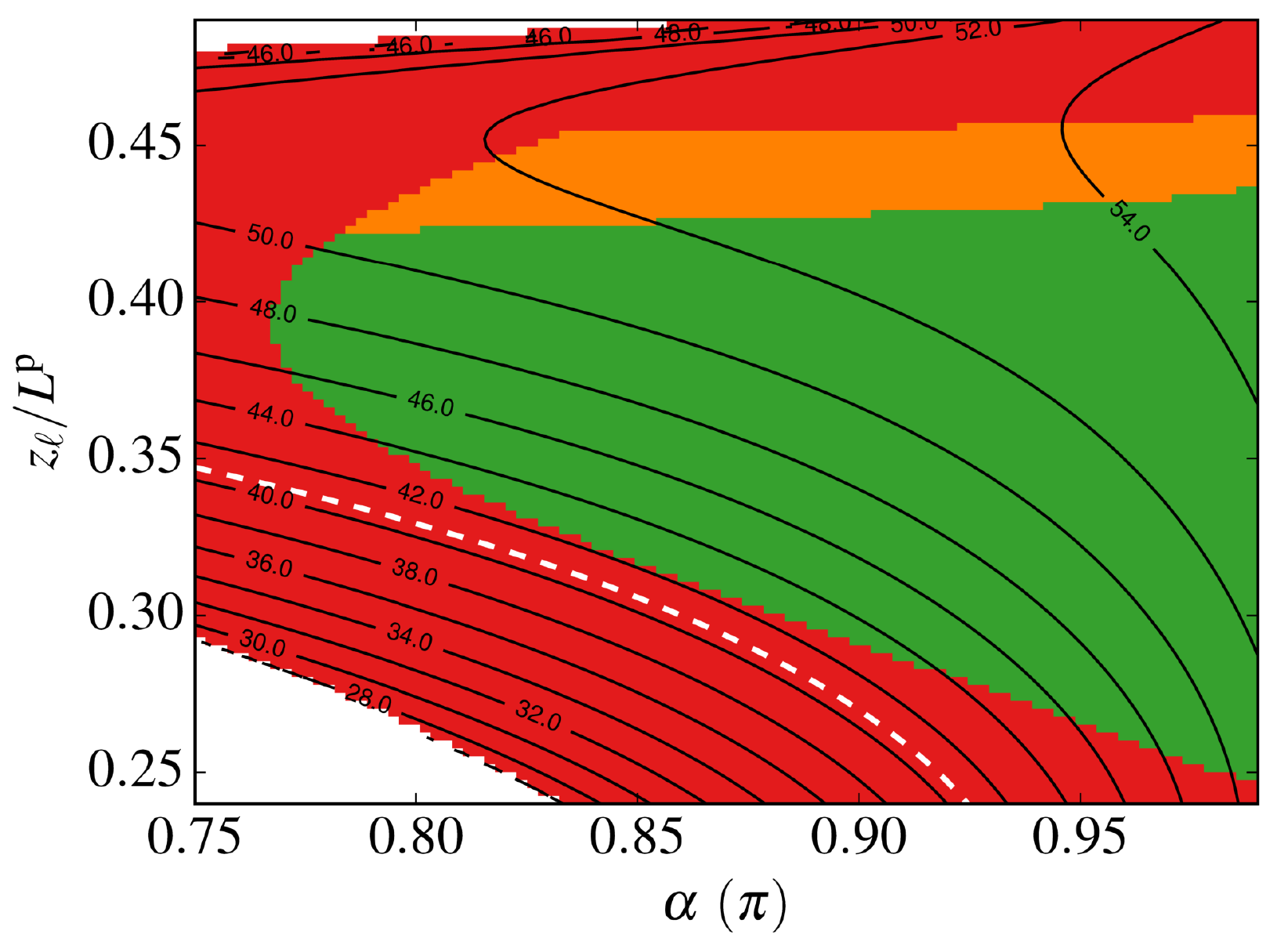}
    \caption{Gas films in closed pore of \SI{0.01}{\micro\meter}}
    \label{fig:gas_film_closed_10nm}
  \end{subfigure}
  \caption{Stability maps of (left column) liquid films and (right
    column) gas films in open and closed pores of lengths
    \SI{10}{\micro\meter} and \SI{0.01}{\micro\meter}. For the closed
    pores, the black contour lines indicate the total fluid density in
    \si{\mol\per\litre}. Further, the gas spinodal density of
    \SI{1.05}{\mol\per\litre} is drawn as a dotted white line and the
    liquid spinodal of \SI{40.7}{\mol\per\litre} is shown as a dashed
    white line. No homogeneous phase filling the entire pore can exist
    for densities between the spinodals. For the open pores, the black
    contour lines indicate the gas phase pressure in
    \si{\kilo\pascal}. Further, the bulk saturation pressure of
    \SI{48.86}{\kilo\pascal} is shown as a dashed white line. In all
    maps, unstable configurations are red, stable configurations are
    green and configurations that are locally stable, but have a
    larger energy than if the pore were filled with a single
    phase, are orange.  }
  \label{fig:film_gas_liquid}
\end{figure*}

The thermodynamic stability of liquid and gas films in open and closed
pores is shown in Figure~\ref{fig:film_gas_liquid}. Configurations are
here mapped in terms of the position of the left contact line $\posl$
and contact angle.

In the open pores, most of the liquid
(Figure~\ref{fig:liquid_film_open_10um}) and gas
(Figure~\ref{fig:gas_film_open_10um}) film configurations are
unstable. The exception is when the film phase is strongly wetting. It
is then possible to have a metastable film, which is evident by the
orange regions in the left part of
Figure~\ref{fig:liquid_film_open_10um} and the right part of
Figure~\ref{fig:gas_film_open_10um}. The stability map for the small
pore in the open system is nearly identical to that of the large pore
and is shown in the supplementary material.

Stability maps of liquid and gas films in the \SI{10}{\micro\meter}
closed pore are shown in Figure~\ref{fig:liquid_film_closed_10um} and
Figure~\ref{fig:gas_film_closed_10um}, respectively. They reveal large
regions where both gas and liquid films are more stable than the
homogeneous phase. In contrast to the adsorbed droplets, which appear
when the liquid is non-wetting, the liquid films are stable only when
the liquid is wetting. Analogously, the gas films in closed pores are
locally stable only for a highly non-wetting liquid.

\begin{figure*}[htbp]
  \centering
  \begin{subfigure}[b]{0.48\textwidth}
    \centering
    \includegraphics[width=\textwidth]{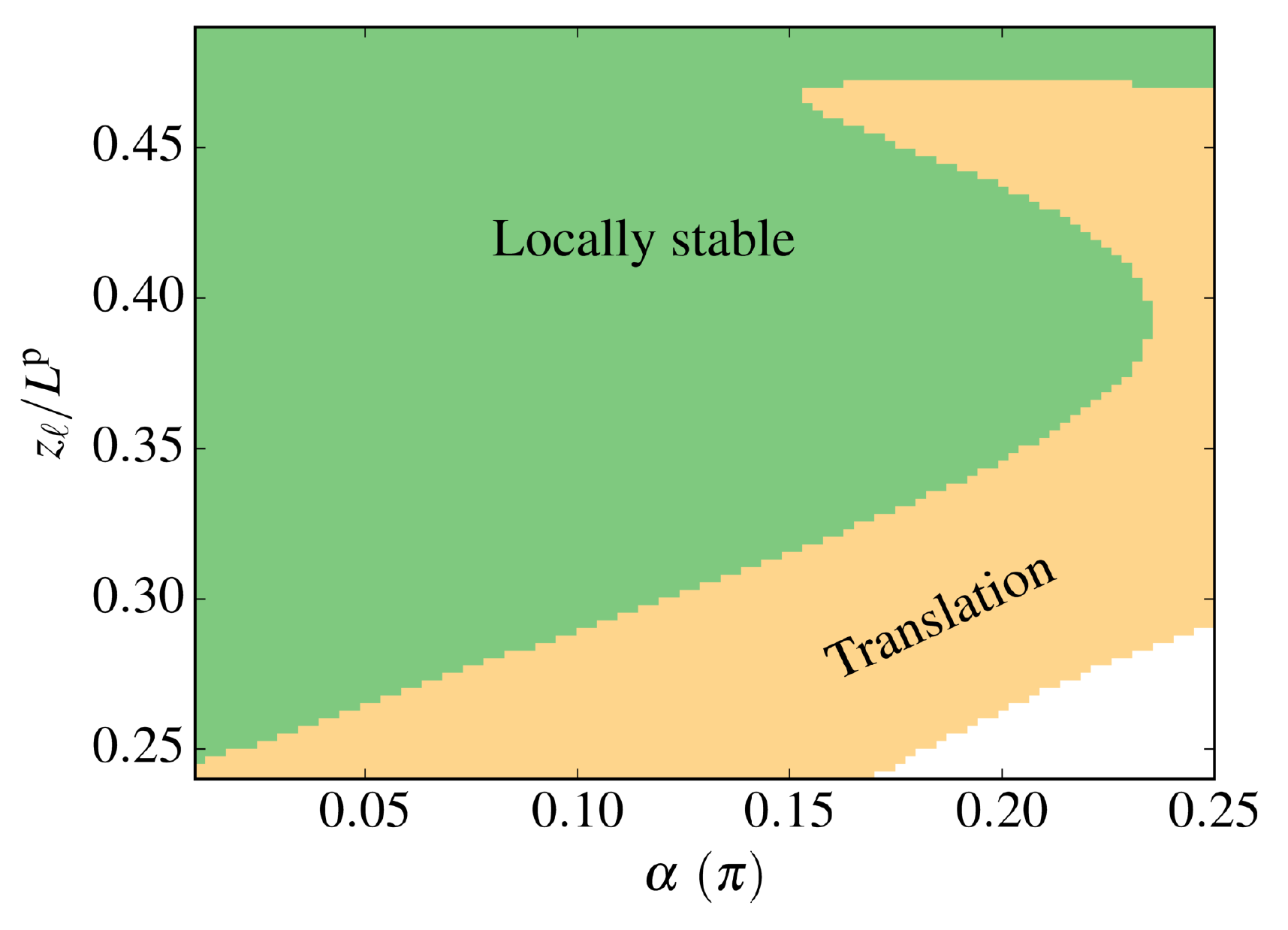}
    \caption{}
    \label{fig:insts_liquid_film_closed_10um}
  \end{subfigure}~
    \begin{subfigure}[b]{0.48\textwidth}
    \centering
    \includegraphics[width=\textwidth]{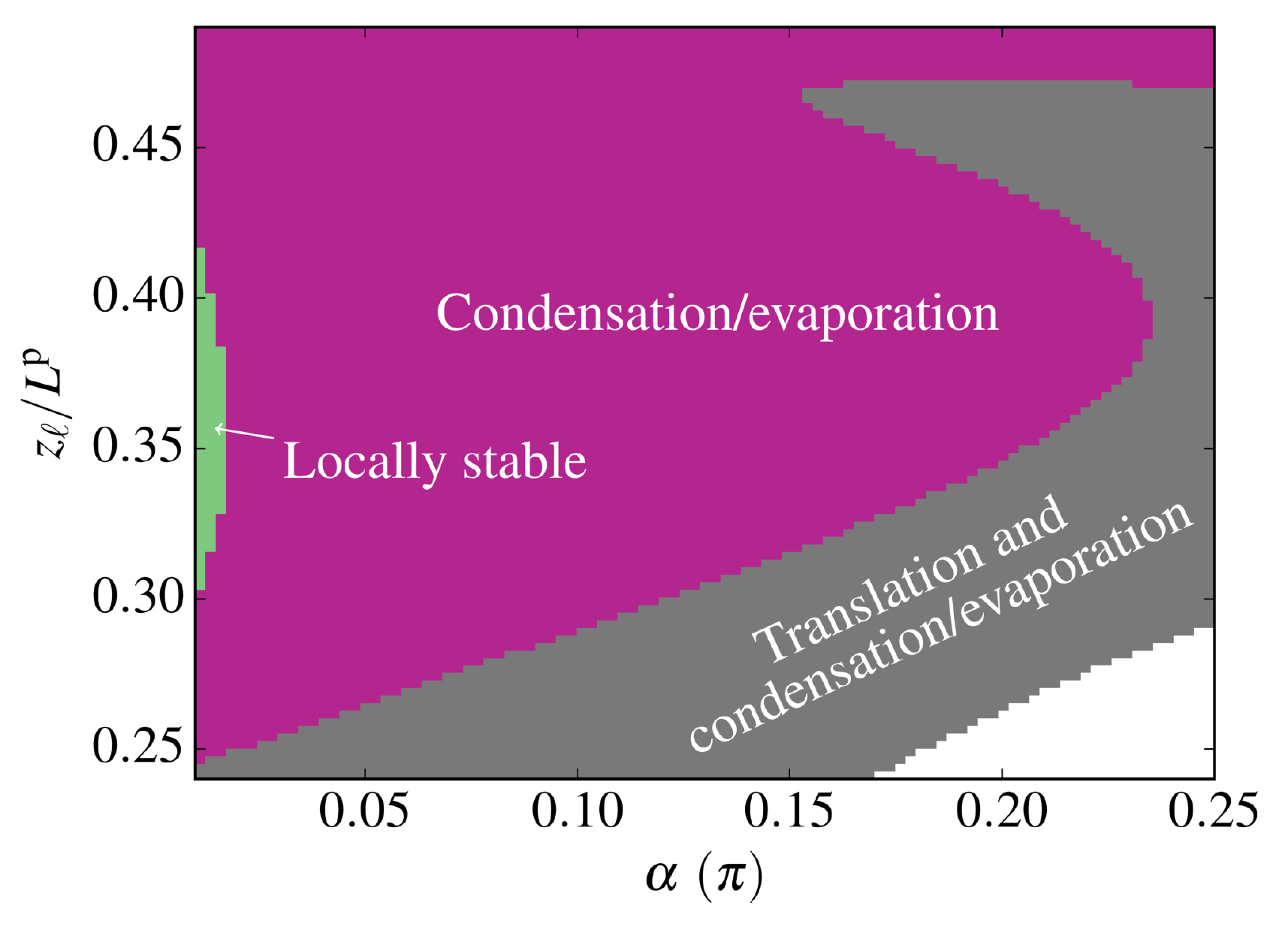}
    \caption{}
    \label{fig:insts_liquid_film_open_10um}
  \end{subfigure}
  \caption{Maps of instability types for liquid films in (a) the
    \SI{10}{\micro\meter} closed pore and (b) the
    \SI{10}{\micro\meter} open pore. Regions with no instabilities are
    light green, regions with only translation instabilities are
    cream, regions with only condensation/evaporation instabilities are
    purple and regions with both types of instabilities are gray.}
\end{figure*}

Figure~\ref{fig:film_gas_liquid} reveals that the thermodynamic
stability of thick films is remarkably different in open and closed
pores. A deeper insight into the origin of this difference can be
gained by further investigating the unstable regions. Like the
adsorbed bubbles and droplets discussed in
Section~\ref{sec:results_adsorbed_bubbledroplet}, a translational
instability is responsible for the unstable regions in the closed
system, as can be seen in
Figure~\ref{fig:insts_liquid_film_closed_10um} for the liquid
film. This translation instability is also present in the same
regions in the open pore (Figure~\ref{fig:liquid_film_open_10um} and
Figure~\ref{fig:gas_film_open_10um}). However, the open pore also has
a condensation/evaporation instability which is present in the entire
unstable region. This is illustrated in
Figure~\ref{fig:insts_liquid_film_open_10um} for the liquid film.

As for the previous two configuration types, a metastable region
appears in the closed system when the pore size is reduced, as shown
in Figure~\ref{fig:liquid_film_closed_10nm} and
Figure~\ref{fig:gas_film_closed_10nm} for the liquid and gas films,
respectively. The metastable configurations represent films whose
small gas and liquid volumes, again, do not compensate for their
interfacial energy costs. In addition, an unstable region also appears
for films with small extent along the $z$-axis. This region is present
for both, but is larger for the gas film than the liquid film. It is
caused by a condensation/evaporation instability.

We emphasize that the appearance of metastable regions and of
condensation/evaporation instabilities cannot be predicted form a
purely mechanical analysis of the film, and a complete thermodynamic
stability analysis is needed. This is important critique to nearly all
previous works that evaluate the stability of films in the literature,
and an important future work is thus to extend the present analysis to
films that have a non-zero disjoining pressure to shed new light on
the thermodynamic stability of thin films in open and closed systems.

\subsection{Phase diagrams}
\begin{figure*}[htbp]
  \centering
  \begin{subfigure}[b]{0.48\textwidth}
    \centering
    \includegraphics[width=\textwidth]{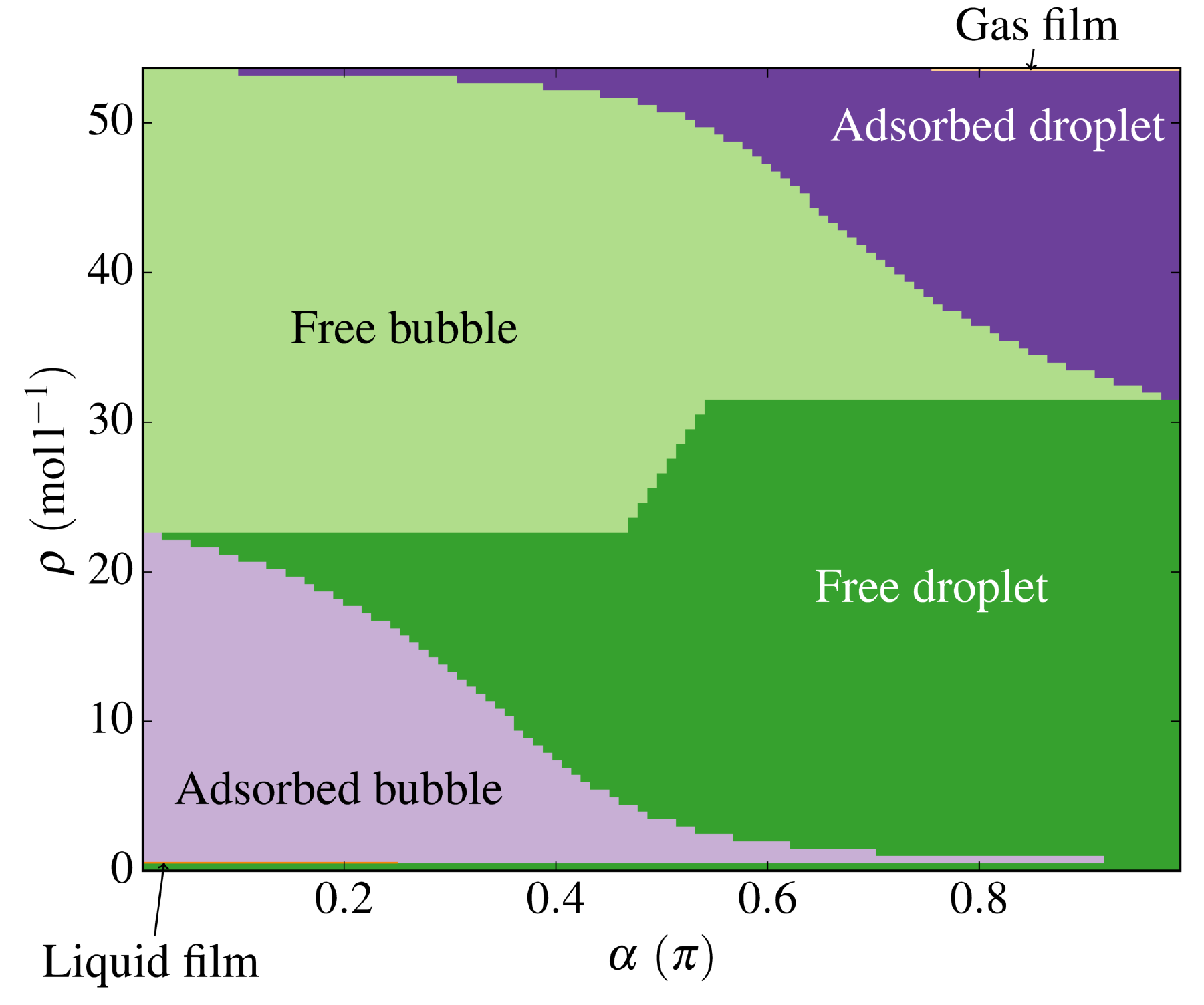}
    \caption{}
    \label{fig:phase_diagram_closed_10um}
  \end{subfigure}~  
  \begin{subfigure}[b]{0.48\textwidth}
    \centering
    \includegraphics[width=\textwidth]{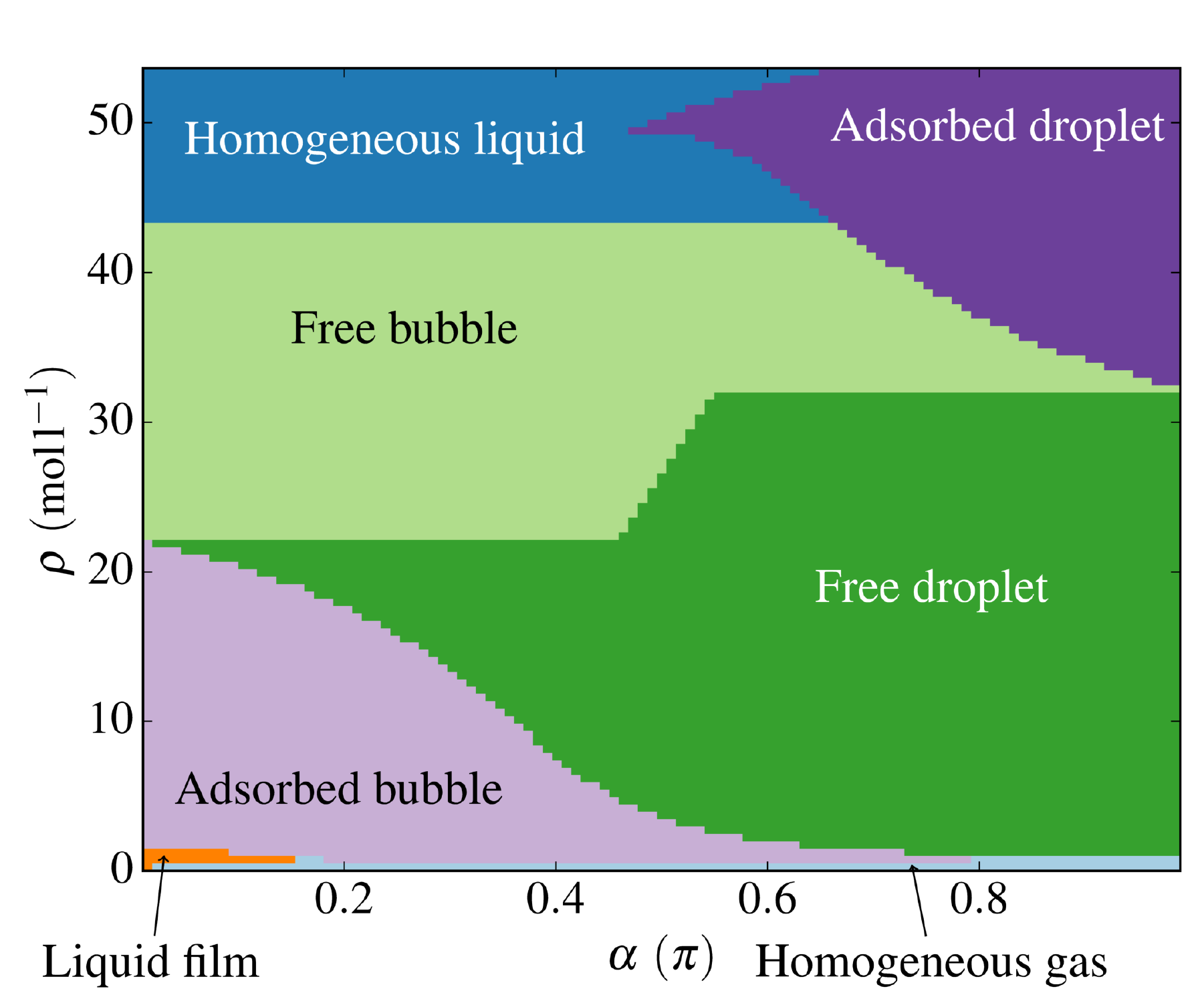}
    \caption{}
    \label{fig:phase_diagram_closed_10nm}
  \end{subfigure}
  \caption{Phase diagram showing equilibrium configurations in a
    closed pores of size (a) $\SI{10}{\micro\meter}$ and (b)
    $\SI{0.01}{\micro\meter}$. Considered configurations are
    homogeneous gas (light blue), homogeneous liquid (blue), free
    bubble (light green), free droplet (green), adsorbed bubble (light
    purple), adsorbed droplet (purple), gas film (light orange) and
    liquid film (orange).}
  \label{fig:phase_diagram_closed}
\end{figure*}

The final results we report are phase diagrams that show the
equilibrium configuration, i.e.\ that which has the lowest total energy
at given conditions. Such phase diagrams are shown in
Figure~\ref{fig:phase_diagram_closed} for closed pores of sizes
$\SI{10}{\micro\meter}$ and $\SI{0.01}{\micro\meter}$. The equilibrium
configurations herein are determined by comparing the Helmholtz
energies of the different (locally stable) fluid configurations at
each value of $\rho$ and $\alpha$.

The phase diagram for the large pore in
Figure~\ref{fig:phase_diagram_closed_10um} exhibits a large degree of
symmetry around the neutral-wetting case where $\alpha = \pi/2$. As
pore-size is reduced (Figure~\ref{fig:phase_diagram_closed_10um}),
however, this symmetry is broken, partly due the appearance of
homogeneous phases with lower Helmholtz energies than the heterogeneous
structures. We note in particular the appearance of a stretched
homogeneous liquid phase at the expense of the free bubble and adsorbed
droplet configurations. In a similar manner, a compressed
homogeneous gas phase appears as well.

For both phase diagrams, the free bubble and adsorbed droplet
configurations prevail at higher densities, because locally stable
configurations of these kinds allow for large liquid volume
fractions. Similarly, the free droplet and adsorbed bubble are
prevalent at lower densities, as these allow for large gas volume
fractions. At densities around $\SI{25}{\mol\per\litre}$, the
equilibrium configuration is either a free bubble (when the liquid is
wetting) or a free droplet (when the liquid is non-wetting).

Equilibrium liquid film configurations are only observed at low
densities and for wetting liquids. In the
$\SI{10}{\micro\meter}$-pore, the extent of the liquid film region
along the $\rho$-axis is smaller than the resolution used in the
figure. A finer resolution of the low densities, however, reveals that
the region is indeed there. Its position and extent along the
$\alpha$-axis is indicated by a thin solid line in
Figure~\ref{fig:phase_diagram_closed_10um}.

The size of the liquid film region grows when the pore size is reduced
(Figure~\ref{fig:phase_diagram_closed_10nm}). This expansion is
primarily at the expense of the adsorbed bubble region and can be
explained as follows. When the pore size is reduced, the liquid
pressures in the adsorbed bubble configurations are also
reduced. Eventually, these pressures reach the liquid spinodal, where
the liquid phase can no longer exist, and it is no longer possible to
have an adsorbed bubble. This is also evident from comparing the white
region in the lower left corner of
Figure~\ref{fig:adsorbed_bubble_open_10nm} with that in
Figure~\ref{fig:adsorbed_bubble_open_10um}. The liquid film, however,
is stable in this region as it has a lower curvature than the adsorbed
bubble. It therefore appears in the phase diagram as the equilibrium
configuration when it is no longer possible to have an adsorbed
bubble.

Like the liquid film region in the $\SI{10}{\micro\meter}$-pore
(Figure~\ref{fig:phase_diagram_closed_10um}), the extent of the gas
film region along the $\rho$-axis is smaller than the resolution. Its
position is therefore also indicated with a thin solid line. The gas
film appears at high densities, when the liquid is non-wetting. For
the $\SI{10}{\nano\meter}$-pore, however, we do not observe any gas
films at equilibrium. The adsorbed droplet is found to always have a
lower energy than the gas film. This conclusion may well change for
other fluids and pore geometries.


\begin{figure*}[htbp]
  \centering
  \begin{subfigure}[b]{0.48\textwidth}
    \centering
    \includegraphics[width=\textwidth]{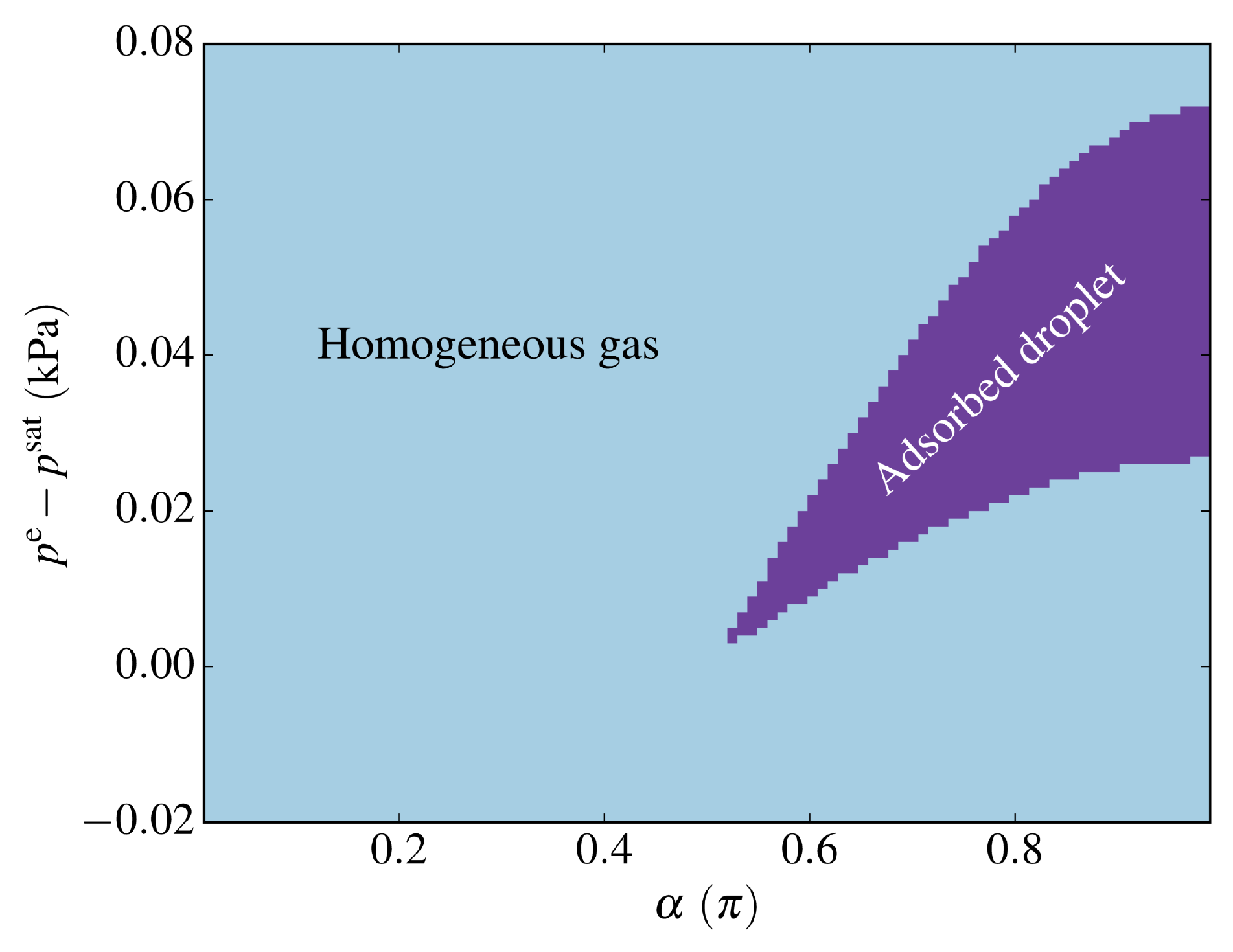}
    \caption{}
    \label{fig:phase_diagram_open_gas_10um}
  \end{subfigure}~  
  \begin{subfigure}[b]{0.468\textwidth}
    \centering
    \includegraphics[width=\textwidth]{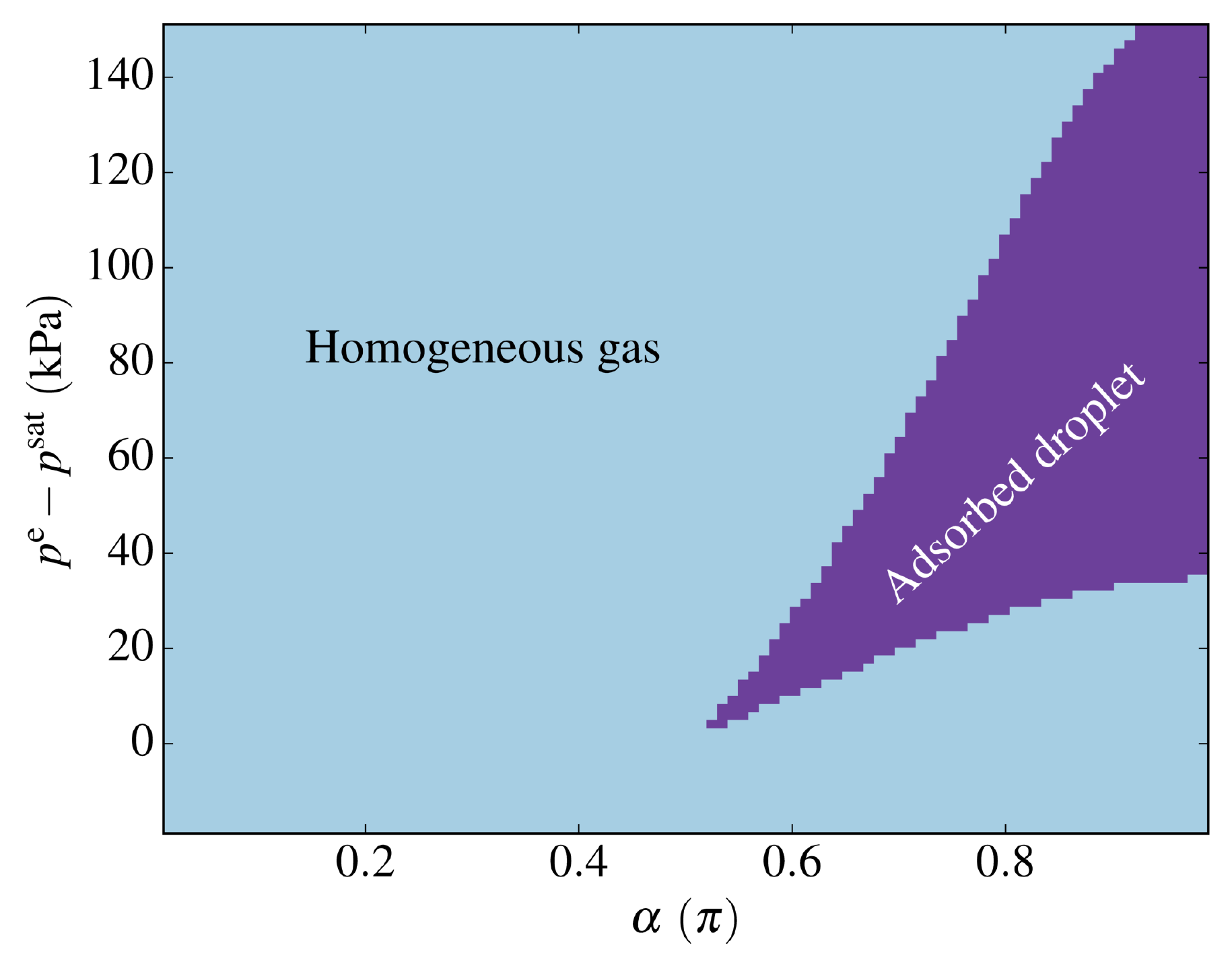}
    \caption{}
    \label{fig:phase_diagram_open_gas_10nm}
  \end{subfigure}
  \caption{Phase diagram showing equilibrium configurations in an open
    pore of size (a) $\SI{10}{\micro\meter}$ and (b)
    $\SI{0.01}{\micro\meter}$ and a gaseous external phase. Considered
    configurations are homogeneous gas (light blue) and adsorbed
    droplet (purple). External pressure minus the bulk saturation
    pressure, $p^{\text{sat}} = \SI{48.86}{\kilo\pascal}$, is
    indicated on the ordinate axis.}
  \label{fig:phase_diagram_open_gas}
\end{figure*}

\begin{figure*}[htbp]
  \centering
  \begin{subfigure}[b]{0.4555\textwidth}
    \centering
    \includegraphics[width=\textwidth]{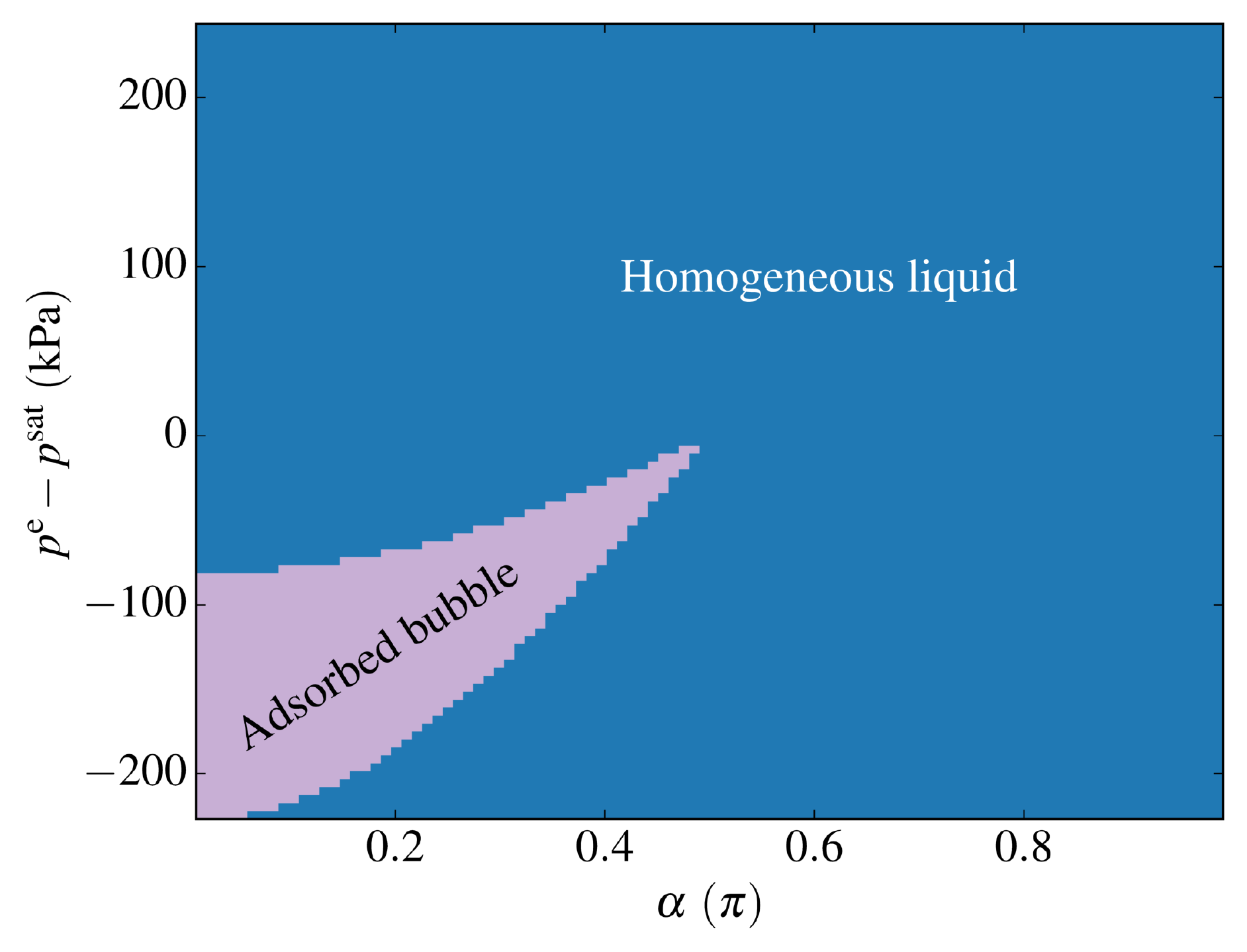}
    \caption{}
    \label{fig:phase_diagram_open_liquid_10um}
  \end{subfigure}~
  \begin{subfigure}[b]{0.48\textwidth}
    \centering
    \includegraphics[width=\textwidth]{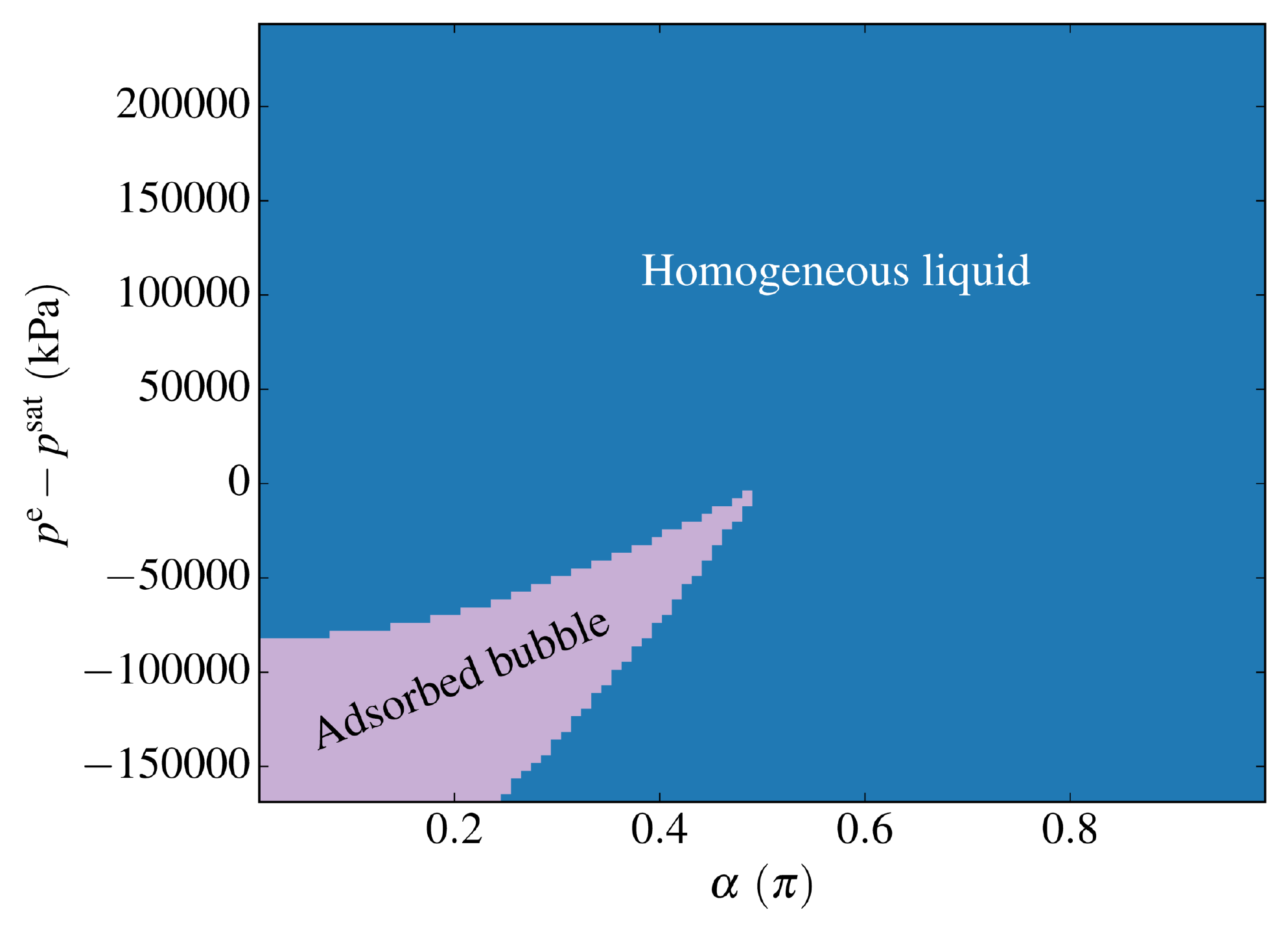}
    \caption{}
    \label{fig:phase_diagram_open_liquid_10nm}
  \end{subfigure}
  \caption{Phase diagram showing equilibrium configurations in an open
    pore of size (a) $\SI{10}{\micro\meter}$ and (b)
    $\SI{0.01}{\micro\meter}$ and a liquid external phase. Considered
    configurations are homogeneous liquid (blue) and adsorbed bubble
    (light purple). External pressure minus the bulk saturation
    pressure, $p^{\text{sat}} = \SI{48.86}{\kilo\pascal}$, is
    indicated on the ordinate axis.}
  \label{fig:phase_diagram_open_liquid}
\end{figure*}

For the closed pores at a given contact angle, a control parameter is
the total density, i.e.\ how many particles the pore contains. For the
open pores on the other hand, the particle reservoir connected to be
pore is either gas or liquid and not both at the same time. We
therefore present separate phase diagrams for cases where the external
phase is gas (Figure~\ref{fig:phase_diagram_open_gas}) and cases when
it is liquid (Figure~\ref{fig:phase_diagram_open_liquid}). The
equilibrium configurations are here determined by comparing the grand
canonical energies of the different (locally stable) fluid
configurations at each value of $p^\ext$ and $\alpha$.

For both pore sizes and external phase choices, a homogeneous phase is
the prevailing fluid configuration. With an external gas phase, an
adsorbed droplet appears as the liquid is non-wetting and, with an
external liquid phase, an adsorbed bubble appears when the liquid is
wetting. Compared to the phase diagrams for the closed pores, there
are much fewer possible configurations.
The free droplets and bubbles are absent
because they are unstable for all possible contact angles and external
pressures. The film configurations are absent because, even though they
may be locally stable in an open pore, the homogeneous phase has a lower
grand canonical energy.

As the pore size is decreased, the range of external pressures at
which we find an adsorbed droplet or bubble is greatly increased. This
is evident by comparing the scale of the ordinates of
Figures~\ref{fig:phase_diagram_open_gas} and
\ref{fig:phase_diagram_open_liquid}. The increase in range is due to
the increased Young--Laplace pressure difference of the interface
induced by the large interfacial curvatures in the small pores.

\section{Conclusion}
\label{sec:conclusion}

We have studied the thermodynamic stability of free and adsorbed
droplets, bubbles and gas and liquid films in open and closed pores by
use of capillary models coupled to an equation of state. We used water
at \SI{358}{\kelvin}, as described by the SRK-CPA equation of state,
as example. Emphasis was placed on the effect of fluid-solid
interaction, as described by a finite contact angle, pore size and
whether the pore is open or closed.

For free droplets and bubbles, our findings were in agreement with
previous works \cite{Wilhelmsen_2014,Yang1985}. These configurations
were unstable in open pores but could be stable in closed pores, if the
bubbles/droplets were large enough.

In contrast to the free droplets and bubbles, adsorbed droplets and
bubbles could be stable both in closed and open pores. Evidently, the
interaction with the solid phase made these structures stable and in
many cases, depending on contact angle and phase fractions, favorable
w.r.t.\ a homogeneous fluid phase.

A new methodology was presented to analyze the thermodynamic stability
of films, where the integral that describes the total energy of the
system was approximated by a quadrature rule. The methodology allowed
us to examine in detail the perturbation that made the film unstable.

Gas and liquid films were found to be unstable at most conditions in
the open pores. The exception was when the fluid was nearly perfectly
wetting and a liquid film could form, or nearly perfectly non-wetting,
when a gas film could form. In both cases, the films were metastable
with respect to the homogeneous phase. In the closed pore, both stable
gas and liquid films were found. As for the free droplets and bubbles,
metastable regions where a homogeneous phase was energetically
preferable were observed for both adsorbed droplets and bubbles and
for the gas and liquid films in the small pore. The reason was that,
as the volumes of the gas and liquid phases became smaller, the energy
gained from having both a liquid and a gas phase did not compensate
for the energy cost of the gas-liquid interface.

Observed instabilities for the adsorbed droplets/bubbles and the films
belonged to one of two distinct classes: (1) translation and (2)
condensation/evaporation. Although exceptions were present, the
general trend was that translation instabilities were observed in the
closed pores while both translation and condensation/evaporation
instabilities were observed in the open pores.

Finally, we presented phase diagrams showing equilibrium configuration
types for both open and closed pores. The closed-pore phase diagrams
were found to contain a larger variety of structures compared to the
open-pore diagrams. Partly, this is because the open-pore diagrams can
contain only structures where the external phase is gas or liquid,
while the closed-pore diagrams can have both kinds of structures.
Most interesting, however, is the lack of locally stable configurations
of free droplets/bubbles and films with lower energy than a homogeneous
phase in the open pores.

The appearance of metastable regions and of condensation/evaporation
instabilities cannot be predicted form a purely mechanical analysis of
the systems. A complete thermodynamic stability analysis, as performed
herein, is necessary. In previous literature on films, the discussion
is usually limited to mechanical stability. The methodology presented
in this work can be used to shed new light on the topic.

The analysis presented in this work is a step towards
developing a thermodynamic framework to map the rich heterogeneous
phase diagram of porous media and other confined systems.



\section*{Acknowledgments}
This work was supported by the Research Council of Norway through its
Centres of Excellence funding scheme, project number 262644.

\section*{Declaration of interest}
Declarations of interest: none.


\appendix

\section{Solutions to the film Euler--Lagrange equation}
\label{app:film_modes}

\begin{figure}[tbp]
  \centering
  \begin{subfigure}[b]{0.48\textwidth}
    \centering
    \includegraphics[width=\textwidth]{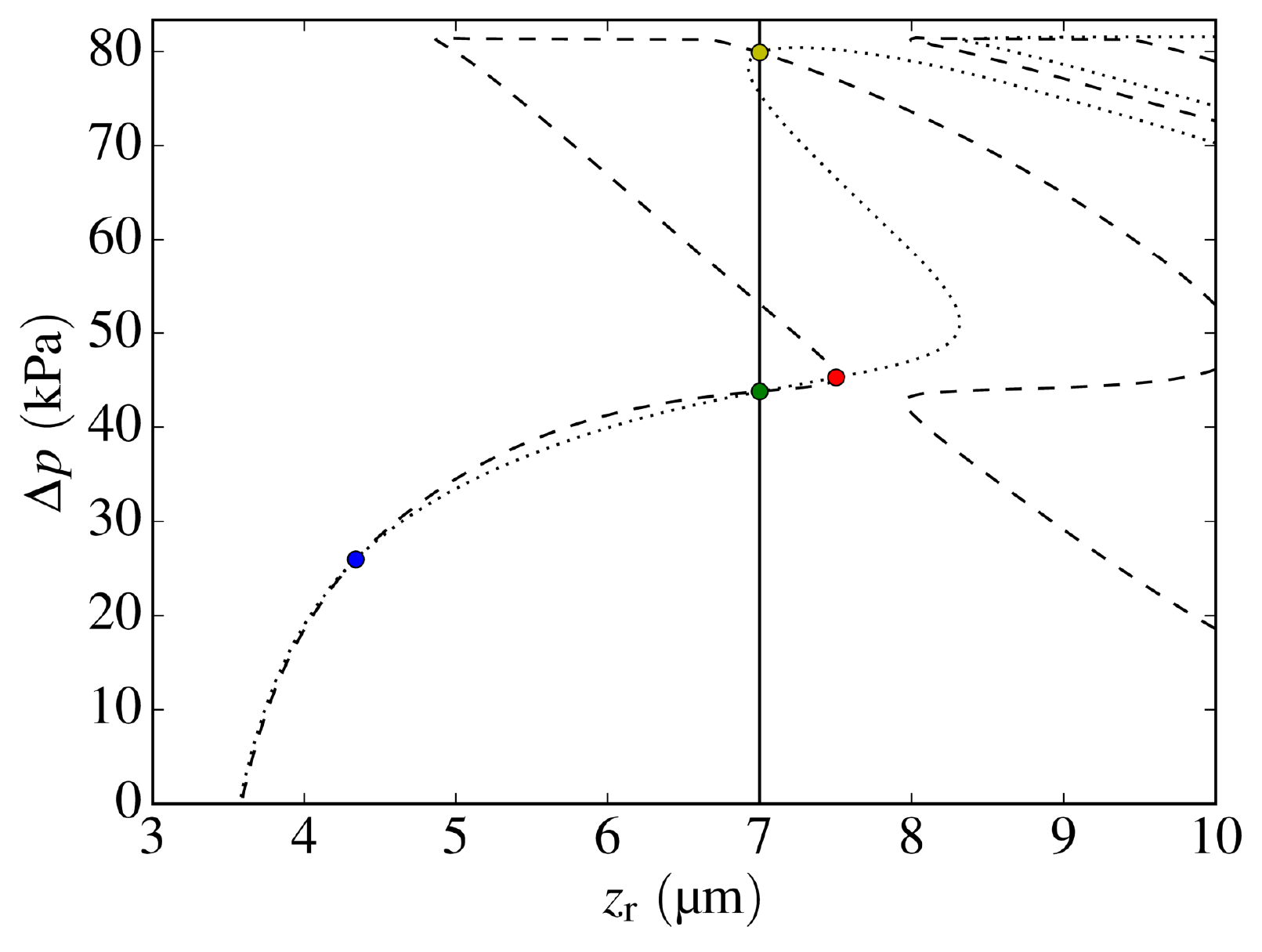}
    \caption{}
    \label{fig:film_modes_map}
  \end{subfigure}
  \begin{subfigure}[b]{0.48\textwidth}
    \centering
    \includegraphics[width=\textwidth]{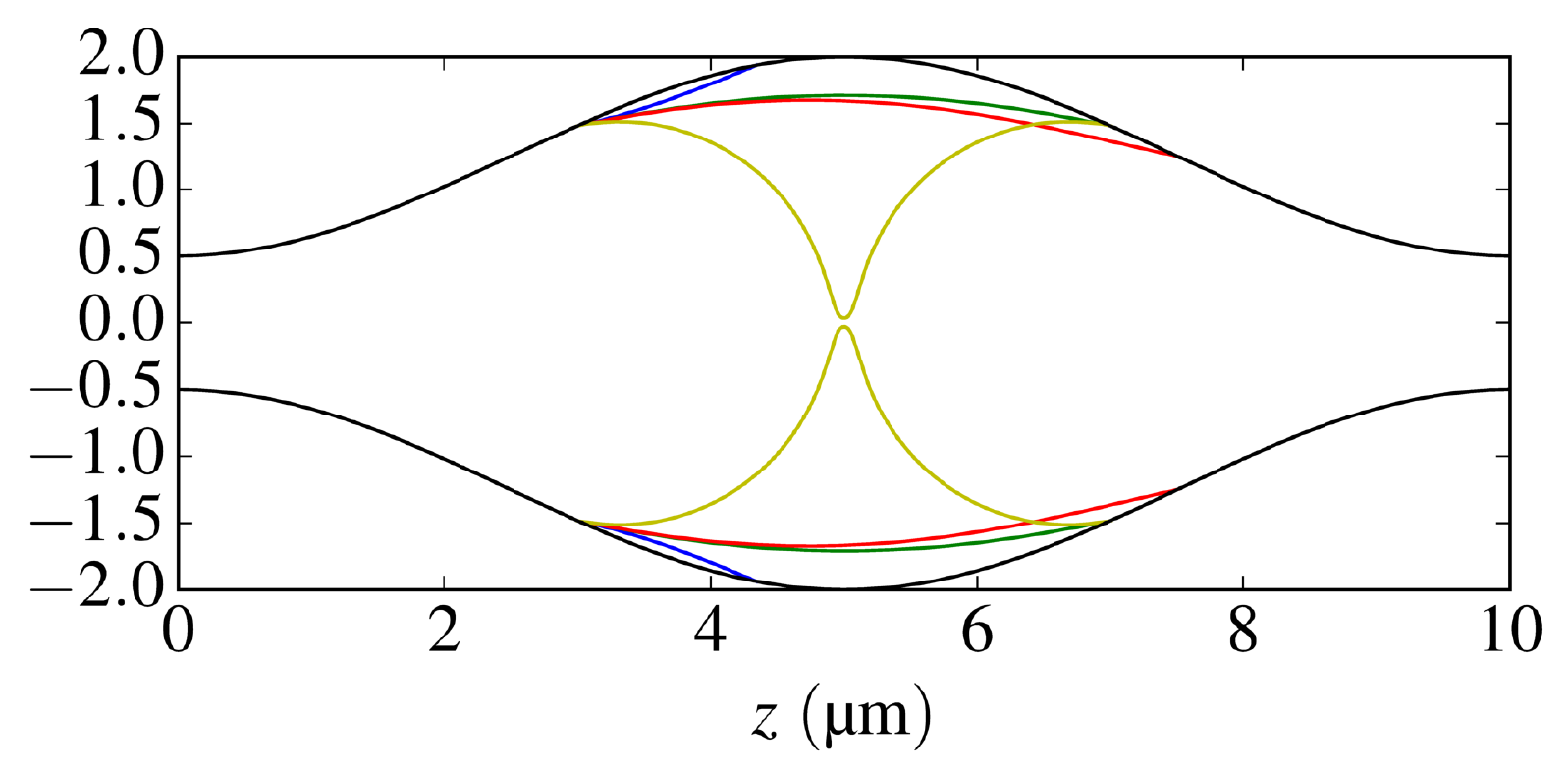}
    \caption{}
    \label{fig:film_modes_geometry}
  \end{subfigure}
  \caption{Illustration of different solutions to the film ODE
    \eqref{eq:film_ode} for a given pore, shown in (b),
    $\posl=\SI{3}{\micro\meter}$ and $\alpha^\nuc = 0.07\pi$. A map of
    the space searched for solutions to the two-point boundary value
    problem $\vec{G} = \vec{0}$ is shown in (a). The dotted curve
    indicates where the first element of $\vec{G}$ is zero and the
    boundary condition on $R^\film$ is satisfied. The dashed curve
    indicates where the second element of $\vec{G}$ is zero and the
    boundary condition on $\dot{R}^\film$ is satisfied. A solution to
    the two-point boundary value problem is thus a point where the two
    curves intersect. Four different solutions are indicated (blue,
    green, red and yellow). The corresponding film profiles are shown
    in (b).}
  \label{fig:film_modes}
\end{figure}

A complicating factor in the search for stationary states for the film
is that there may be many solutions to $\vec{G} = \vec{0}$ with
a specified $\posl$. $\vec{G}$ is the residual function for the
two-point boundary value problem obtained by setting
\eqref{eq:shooting_residuals} equal to $\vec{0}$. One example is
illustrated in Figure~\ref{fig:film_modes}, which displays a pore with
$L^\pore = \SI{10}{\micro\meter}$, where the shape of the pore is
defined by \eqref{eq:R_p} and drawn in black in
Figure~\ref{fig:film_modes_geometry}. Furthermore, the contact angle
is $\alpha^\nuc = 0.07\pi$, $\sigma^{\ext\nuc} =
\SI{0.0616}{\newton\per\meter}$ and $\posl = \SI{3}{\micro\meter}$.
Figure~\ref{fig:film_modes_map} maps the search space. The dotted
curve shows where the first element of $\vec{G}$ is zero and the
boundary condition on $R^\film$ is satisfied, while the dashed curve
shows where the second element of $\vec{G}$ is zero and the boundary
condition on $\dot{R}^\film$ is satisfied. A solution to the two-point
boundary value problem is thus a point where the two curves
intersect. The solid vertical line is drawn at $\posr=L^\pore -
\posl$. Solutions falling on this line are symmetric or anti-symmetric
with respect to the center of the pore.

Four different solutions are indicated by circles (blue, green, red
and yellow) in Figure~\ref{fig:film_modes_map}. The corresponding film
configurations are shown in Figure~\ref{fig:film_modes_geometry}. Of
these four solutions, only the green and the yellow are
symmetric. When evaluating the thermodynamic stability of these
solutions, however, it turns out that only the green solution is
stable in a closed system. All solutions are unstable in an open
system. For a more thorough discussion on thermodynamic stability, we
refer to Section~\ref{sec:results}. Since we observe that solutions
that are not symmetric around the pore center are always unstable, we
only need to consider the symmetric film solution with the lowest
$\Delta p$, that is also feasible in the sense that $0 < R^\film
\left(z\right) < R^\pore \left(z\right) \forall z \in \left( \posl,
\posr \right)$. In the analysis presented in this paper, we ignore the
other solutions, even if they also represent unstable stationary
states of the film. These unstable states may, however, be of interest
in the theory of nucleation. For instance, the yellow profile in
Figure~\ref{fig:film_modes_geometry} could well be the saddle point
that determines the activation barrier to the creation of a an
adsorbed droplet, similar to the one depicted in
Figure~\ref{fig:pore_with_adsorbed_droplet}.

\section{Convergence of the discrete film description method}
\label{app:film_convergence}
In Section~\ref{sec:film_discr}, we presented a discrete method to
describe the Helmholtz energy of the film. Here, we perform a
convergence study to show that the solutions provided by the discrete
method converge to those obtained by solving the Euler--Lagrange
equations when the discrete grid is refined. To this end, we consider
a pore described by \eqref{eq:R_p} with $L^\pore =
\SI{10}{\micro\meter}$ and choose $\alpha^\nuc = \pi/20$ and
$\sigma^{\ext\nuc} = \SI{0.02}{\newton\per\meter}$. The fluid is, as
in Section~\ref{sec:results}, water at \SI{358}{\kelvin}, described by
the CPA-SRK EOS. The film is liquid and the surrounding phase gas.

We use the variational formulation to obtain a stationary state of $F$
where the film starts at $\posl = 0.3 L^\pore$. This solution will
serve both as a reference solution and to generate initial guesses for
the discrete solutions. Subsequently, we solve \eqref{eq:dFdy_0_film}
for different number of grid points, $M$. Relative errors in the film
profile $\vec{R}^\film$ with respect to the reference solution, as
measured in the $L_2$- and $L_\infty$-norms, and the corresponding
estimated convergence orders are presented in
Table~\ref{tbl:R_f_errors}. It is clear from these results that the
discrete solutions converge to the variational solution as the grid is
refined, and that the convergence is second-order.

\begin{table}[htbp]
  \caption{Relative errors in the film profile $\vec{R}^\film$ with
    respect to the reference solution, as measured in the $L_2$- and
    $L_\infty$-norms, for different discrete grid sizes $M$. The
    corresponding estimated convergence orders are also given.}
  \label{tbl:R_f_errors}
  \centering 
  \begin{tabular}{l l l l l} 
    \toprule 
    $M$ & $L_2$-error & $L_2$-order & $L_\infty$-error & $L_\infty$-order \\ 
    \midrule 
    25 & \SI{5.10E-05}{} & - & \SI{8.58E-04}{} & - \\ 
    50 & \SI{6.91E-06}{} & \SI{2.88}{} & \SI{1.99E-04}{} & \SI{2.11}{} \\ 
    100 & \SI{1.00E-06}{} & \SI{2.79}{} & \SI{4.74E-05}{} & \SI{2.07}{} \\ 
    200 & \SI{1.56E-07}{} & \SI{2.69}{} & \SI{1.15E-05}{} & \SI{2.04}{} \\ 
    400 & \SI{2.55E-08}{} & \SI{2.61}{} & \SI{2.84E-06}{} & \SI{2.02}{} \\ 
    \bottomrule 
  \end{tabular}
\end{table}

\end{document}